\documentclass[aps,twocolumn,epsfig,graphics,showpacs,floatfix,mathbbm,noshowpacs]{revtex4}
\usepackage{amsmath,amsfonts,amssymb,graphics,graphicx,epsfig,color,times,bbm}

\newcommand{\id}{{\sf 1 \hspace{-0.3ex} \rule[0.01ex]{0.12ex}{1.52ex} \rule[.01ex]{0.3ex}{0.12ex} }}

\begin{document}

\title{Dynamics 
and manipulation of entanglement in  \\
coupled harmonic systems
with many degrees of freedom}
\author{M.B.\ Plenio$^1$, J.\ Hartley$^1$, and J.\ Eisert$^{2,1}$}
\address{1 Blackett Laboratory, Imperial College London,
Prince Consort Rd, London SW7 2BW, UK\\
2  Institut f{\"u}r Physik, Universit{\"a}t Potsdam,
Am Neuen Palais 10, D-14469 Potsdam, Germany}
\date{\today}

\begin{abstract}
We study the entanglement dynamics of a system consisting of a
large number of coupled harmonic oscillators in various
configurations and for different types of nearest neighbour
interactions. For a one-dimensional chain we provide compact
analytical solutions and approximations to the dynamical evolution
of the entanglement between spatially separated oscillators. Key
properties such as the speed of entanglement propagation, the
maximum amount of transferred entanglement and the efficiency for
the entanglement transfer are computed. For harmonic oscillators
coupled by springs, corresponding to a phonon model, we observe a
non-monotonic transfer efficiency in the initially prepared amount
of entanglement, i.e., an intermediate amount of initial
entanglement is transferred with the highest efficiency. In
contrast, within the framework of the rotating wave approximation
(as appropriate e.g. in quantum optical settings) one finds a
monotonic behaviour. We also study geometrical configurations that
are analogous to quantum optical devices (such as beamsplitters
and interferometers) and observe characteristic differences when
initially thermal or squeezed states are entering these devices.
We show that these devices may be switched on and off by changing
the properties of an individual oscillator. They may therefore be
used as building blocks of large fixed and pre-fabricated 
but programmable structures
in which quantum information is manipulated through propagation.
We discuss briefly
possible experimental realisations of systems of interacting
harmonic oscillators in which these effects may be confirmed
experimentally.\\

Submitted to New J.\ Phys.\

\end{abstract}

\maketitle


\section{Introduction}
Quantum information processing requires as a basic ingredient the
ability to transfer quantum information between spatially
separated quantum bits, either to implement a joint unitary
transformation or, as a special case, to swap quantum information
between the qubits. For a transfer over larger distances it is
usually imagined that some stationary qubits, for example in the
form of trapped ions inside an optical resonator, are coupled to a
quantized mode of the electro-magnetic field that propagates
between the spatially separated cavities \cite{Cirac ZKM 97,Duan K
03,Feng ZLGX 03,Browne PH 03}. Other specific realisations are
possible, but the basic principle always relies on the use of some
continuous degree of freedom between the qubits and their
manipulation by external fields. While this appears to be the most
realistic mode of transport over long distances, one may conceive
other modes over shorter distances. Instead of using a
quantum field one may study the possibilities offered by a
discrete set of interacting quantum systems. This might involve
spin degrees of freedom \cite{Khaneja G 02,Bose 02,Subrahmanyan
03,Christandl DEL 03,Osborne L 03} or infinite dimensional systems
such as harmonic oscillators \cite{Eisert P 03,Eisert PBH 03}. In
the present paper we explore the dynamics of entanglement in a
chain of coupled harmonic oscillators \cite{Audenaert EPW 02,Brun,Halliwell}. 
Apart from its obvious relevance to
quantum optical systems including photonic crystals, such a model
also describes phonons in a crystal and we therefore hope that the
results presented here will also have applications in condensed
matter systems as well.  

This paper is organised as follows. In Section 2 we present the
basic physical models and their Hamiltonians. We use 
analytical tools from the theory of Gaussian states in continuous
variable systems where some rapid development has been achieved
recently (for a tutorial overview see, e.g., Ref.\ \cite{Eisert P 03}). 
We briefly reiterate those results that will be employed in
the present investigation. 
The entanglement
properties of a system of harmonic oscillators in the static
regime have been studied in some detail \cite{Audenaert EPW 02}.
Section 3 will then present the basic
equations of motion in compact form for two types of interactions,
namely (a) harmonic oscillators coupled by springs and, resulting
from this, (b) a model which corresponds to a rotating wave
approximation as is appropriate in a quantum optical setting.
Section 4 employs these equations for the propagation of
entanglement along a chain of harmonic oscillators which might be
realized by coupled nano-mechanical oscillators 
\cite{Roukes 01,Schwab HWR 00} or optical cavities.  The time evolution of the
entanglement between a pair of oscillators is given analytically
in a compact form. Properties such as the speed of propagation,
the amount of entanglement and the transfer efficiency are then
obtained from these expressions. In Section 5 we present results
utilizing the equations from Sections 3 and 4. Firstly, we study a
method for the creation of entanglement in such a system that does
not require detailed control of the interaction strength between
individual oscillators but only the ability for changing the
interaction strength globally \cite{Eisert PBH 03}. The influence
of imperfections such as finite temperatures or randomly varying
coupling constants on such a scheme are studied. We also consider
the propagation of some initially prepared entangled state along
the chain. Surprisingly, for harmonic oscillators coupled by
springs we observe a non-monotonic transfer efficiency in the
initially prepared amount of entanglement, i.e., an intermediate
amount of entanglement is transferred with the highest efficiency.
Conversely, in the rotating wave approximation, the transfer
efficiency is monotonic. While most of these results assume a
position independent and stationary coupling we also show that
with carefully chosen position dependent coupling the transfer
efficiency in this system may be increased to unity. Finally we
study geometrical configurations that are analogous to quantum
optical devices such as beamsplitters and interferometers and
observe characteristic differences when initially thermal or
squeezed states are entering these devices. We show that these
devices may be switched on and off by changing the properties of
an individual oscillator and may therefore be building blocks of
large fixed but programmable structures. In Section 6 we summarize
the results of this paper and suggest possible experimental
realisations of systems of harmonic oscillators in which these
effects may be confirmed.

\section{Models and Methods}
In this section we present the systems under consideration, namely
coupled harmonic oscillators, together with the Hamiltonians that
describe the various models for their interaction. We will
restrict our attention to Hamiltonians that are quadratic in
position and momentum operators. This will be crucial for the
following analysis as it permits us to draw on the results and
techniques from the theory of Gaussian continuous variable
entanglement. The most important results from this theory will be
reviewed here briefly.

\subsection{The physical models}
The general setup consists of a chain of $M$ coupled harmonic
oscillators, where the coupling is assumed to be such that the
corresponding Hamiltonian is at most quadratic in position and
momentum. We will number the harmonic oscillators from $1$ to $M$
with periodic boundary conditions such that the $(M+1)$-th
oscillator is identified with the $1$-st. The choice of periodic
boundary conditions yields exact and compact analytical solutions
since we can employ normal coordinates straightforwardly. A
similar approach is less successful in the non-periodic boundary
case. In addition, we allow for the existence of a distinguished
decoupled oscillator with index $0$ which will be a convenient
notation for some of the later studies. Arranging the position and
momentum operators in the form of a vector
\begin{equation}
    R = ({\hat q}_0,{\hat q}_1,\ldots,{\hat q}_M,{\hat p}_0,
    {\hat p}_1,\ldots,{\hat p}_M)
\end{equation}
we can then write the general Hamiltonian in the form
\begin{equation}\label{hamcompact}
    \hat{H}=\frac{1}{2} R \left[ \begin{array}{cc} V & 0\\
    0 & T
    \end{array}\right]R^T
    = \frac{1}{2}\sum_{ij=1}^M {\hat q}_i V_{ij} {\hat q}_j
    + {\hat p}_i T_{ij} {\hat p}_j,
\end{equation}
where $V$ is the potential matrix and $T$ the kinetic matrix. We
will consider three basic settings for which we now provide the
matrices $T$ and $V$ explicitly. In all these cases we assume that
the oscillators in the chain are all identical with a mass $m=1$
and eigenfrequency $\omega=1$.
\begin{itemize}
\item[(a)] (Uncoupled oscillators) If the oscillators are not coupled to
each other, then the potential energy of the $k$-{th} oscillator
is simply given by ${\hat q}_k^2/2$ while its kinetic energy is  $
{\hat p}_k^2/2$. As a consequence, both the potential matrix and
the kinetic matrix are diagonal and identical, namely
$V=T=\id_{M+1}$, where $\id_{M+1}$ denotes the $M+1$ by $M+1$
identity matrix.

\item[(b)] (Oscillators coupled by springs) If neighbouring oscillators
(except for the $0$-{th} oscillator) are coupled via springs that
obey Hooke's law, the Hamiltonian is given by
\begin{equation}\label{hookes}
    \hat{H}_{\rm Spring}= \frac{{\hat q}_0^2 + {\hat p}_0^2}{2} +
    \frac{1}{2}\sum_{k=1}^{M} {\hat q}_k^2 + {\hat p}_k^2 +
    c\left(\hat{q}_{k+1}-\hat{q}_k\right)^2,
\end{equation}
where $c$ denotes the coupling strength and we have used periodic
boundary conditions i.e., $q_{M+1}=q_{1}$. Keeping in mind that we
wish to leave the oscillator with index $0$ uncoupled, the
potential matrix becomes
\begin{equation}
    V=\left[\begin{array}{ccccccc}
    1     &   0  &  0   &  0   &    0   &  0 & 0\\
    0     &1+2c  & -c   &  0   & \cdots & 0  & -c \\
    0     &-c    & 1+2c & -c   &        &    & 0 \\
    0     &0     & -c   & 1+2c & \ddots &    & \vdots  \\
    \vdots&\vdots&      &\ddots& \ddots & -c & 0\\
    0     &  0   &      &      &   -c   &1+2c&-c\\
    0     &  -c  &   0  &\cdots&   0    &  -c&1+2c
\end{array}
\right],
\end{equation}
while the kinetic matrix is given by the identity matrix
$T=\id_{M+1}$.

\item[(c)] (Oscillators in rotating wave approximation)  An interaction
that provides simpler dynamics is obtained via the rotating wave
approximation in quantum optical systems.  Indeed, if we define
annihilation and creation operators
\begin{eqnarray}
    {\hat a} &=& \frac{1}{\sqrt{2}} ( {\hat q} +i{\hat p}),
    \;\;\;\;\;\;
    {\hat a}^{\dagger} = \frac{1}{\sqrt{2}} ( {\hat q} -i{\hat p}),
\end{eqnarray}
then we observe that the interaction in case (b) includes terms of
the form ${\hat a}^{\dagger}_k {\hat a}^{\dagger}_{k+1}$, i.e.,
interaction terms for which both harmonic oscillators are being
excited simultaneously. Such terms are not energy conserving, and
in quantum optics they are usually be neglected in the framework
of the rotating wave approximation (RWA). Following this practice
amounts to considering the following Hamiltonian
\begin{eqnarray}
    \hat{H}_{\rm RWA} & =& {\hat a}^{\dagger}_0 {\hat a}_0 +\frac{1}{2} + \sum_{k=1}^{M}
    (1+c)\Bigl({\hat a}^{\dagger}_k {\hat a}_k + \frac{1}{2}\Bigr)\nonumber\\
    &-& c\left(\hat{a}^{\dagger}_{k+1}\hat{a}_k +
    \hat{a}_{k+1}\hat{a}^{\dagger}_k\right).
\end{eqnarray}
In terms of position and momentum operators this can be written as
\begin{eqnarray}
    \hat{H}_{\rm RWA} & =& \frac{{\hat q}_0^2 + {\hat p}_0^2}{2}+
    \frac{1}{2} \sum_{k=1}^{M}
    {\hat q}_k^2 + {\hat p}_k^2 \nonumber \\ &+&
    \frac{c}{2} \left(\hat{q}_{k+1}-\hat{q}_k\right)^2 +
    \frac{c}{2} \left(\hat{p}_{k+1}-\hat{p}_k\right)^2,\label{rwa}
\end{eqnarray}
so that in this case both the potential and the kinetic matrix are
given by
\begin{equation}
    T=V=\left[\begin{array}{ccccccc}
    1     &   0  &  0   &  0   &    0   &  0 & 0\\
    0     &1+c  & -c/2   &  0   & \cdots & 0  & -c/2 \\
    0     &-c/2    & 1+c & -c/2   &        &    & 0 \\
    0     &0     & -c/2   & 1+c & \ddots &    & \vdots  \\
    \vdots&\vdots&      &\ddots& \ddots & -c/2 & 0\\
    0     &  0   &      &      &   -c/2   &1+c&-c/2\\
    0     &  -c/2  &   0  &\cdots&   0    &  -c/2&1+c
\end{array}
\right].
\end{equation}
\end{itemize}

Note that the matrix $V$ can be conceived as the adjacency matrix
of a weighted graph $G(v,e)$ encoding the interaction pattern
between the systems in the canonical coordinates corresponding to
position. Vertices of the graph correspond to physical systems,
i.e., the individual harmonic oscillators, whereas the weight
associated with each of the edges quantifies the coupling strength
\cite{Wolf}. The main diagonal corresponds to loops of the
weighted graph. This intuition is in immediate analogy to graph
states for spin systems with an Ising interaction between the
constituents \cite{Graphs,Schlinge,Rauss}  and can be useful in
the study of more complex geometries. Before we study the
dynamical properties of these systems, we provide in the following
subsection a brief overview over the main technical tools that we
are going to employ.

\subsection{Analytical tools}
Analysing the entanglement properties of infinite dimensional
systems such as harmonic oscillators is generally technically
involved unless one restricts attention to specific types of
states. 
Indeed, in recent years a detailed
theory of so-called Gaussian entangled states has been developed.
As we will employ some of its basic results in the subsequent
analysis we are providing a brief review of some useful results. A
more detailed tutorial review can be found, e.g., in Ref.\
\cite{Eisert P 03}, and more technical details concerning Gaussian
states can be found in Ref.\ \cite{Simon SM 87}.

The relevant variables in the analysis of harmonic oscillators are the
canonical operators for position and momentum. Let us assume a
system with $n$ harmonic oscillators. As stated above it is
convenient to arrange these in the form of a vector
\begin{displaymath}
    R^T = ({\hat q}_1,\ldots,{\hat q}_n,{\hat p}_1,\ldots,{\hat
    p}_n)\, .
\end{displaymath}
The characteristic feature distinguishing the quantum harmonic
oscillator from its classical counterpart is the canonical
commutation relation (CCR) between position and momentum.
Employing the vector $R$ these can be written in the particularly
convenient form $ [\hat R_j,\hat R_k]=i \sigma_{j,k}$ where the
real skew-symmetric block diagonal $2n\times 2n$-matrix $\sigma$,
the symplectic matrix, given by
\begin{equation}
    \sigma =\left[\begin{array}{cccccc}
    0        & {\id }_n\\
    -{\id }_n &     0    \end{array}
    \right],
\end{equation}
assuming units where $\hbar=1$, and $k=1$, a choice that will be
adopted for the rest of this paper. Instead of referring to
states, i.e., density operators, one may equivalently refer to
functions that are defined on phase space. While there are many
equivalent choices for phase space distributions, for the purposes
of this work it is most convenient to introduce the (Wigner-)
characteristic function. Using the Weyl operator $W_{\xi} = e^{i
\xi^{T} \sigma R}$ for $\xi\in{\mathbf{R}}^{2n}$, we define the
characteristic function as
\begin{equation}
    \chi_{\rho}(\xi) = \mbox{tr}[\rho W_{\xi}].
\end{equation}
The state and its characteristic function are related to each
other according to a  Fourier-Weyl relation,
\begin{equation}
    \rho = \frac{1}{(2\pi)^n} \int d^{2n}\xi
    \chi_{\rho}(-\xi) W_{\xi} .
\end{equation}
Gaussian states are exactly those states for which the
characteristic function $\chi_{\rho}$ is a Gaussian function in
phase space \cite{Simon SM 87}. That is, if the characteristic
function is of the form
\begin{equation}
    \chi_{\rho}(\xi) =
    \chi_{\rho}(0) e^{-\frac{1}{4}(\sigma\xi)^T \gamma (\sigma\xi) - d^T
    (\sigma\xi)}.
\end{equation}
As is well known, Gaussians are completely specified by their
first and second moments, $d$ and $\gamma$ respectively.  As the
first moments   can be always made zero
utilizing appropriate local displacements in phase space, they are
not relevant in the context of questions related to squeezing and
entanglement and will be ignored in the following. The second
moments can be collected in the real symmetric $2n\times 2n$
covariance matrix $\gamma$ defined as
\begin{eqnarray}
    \gamma_{j,k} &=&
    2 \mbox{Re} \, \mbox{tr}\bigl[
    \rho ({\hat R}_j-\langle {\hat R}_j\rangle_{\rho}  )
    ({\hat R}_k-\langle {\hat R}_k\rangle_{\rho}  )
    \bigr] \, .
\end{eqnarray}
With this convention, the covariance matrix of the $n$-mode vacuum
is $\gamma=\id_{2n}$.

As the covariance matrix encodes the complete information about
the entanglement properties of the system, we will use it in order
to quantify the amount of entanglement between two groups of
oscillators. There is no unique way to quantify entanglement for
mixed states, and several different measures grasp entanglement in
terms of different operational interpretations. For the purposes
of this work we settle for the logarithmic negativity which is
comparatively easy to compute and possesses an interpretation as a
cost function \cite{Audenaert PE 03,Lewenstein HSL 98,Vidal W
02,Eisert P 99,Eisert Phd}. Given two parties, $A$ and $B$, the
logarithmic negativity is defined as
\begin{equation}
    N(\rho)=\log_2||\rho^{T_B}||_1
\end{equation}
where $\rho^{T_B}$ is the state that is obtained from $\rho$ via a
partial transposition with respect to system $B$ and $||.||_1$
denotes the trace-norm. As we focus attention on Gaussian states
which we characterize via the covariance matrix $\gamma$ rather
than the density matrix $\rho$, we need to provide a prescription
for the evaluation of the logarithmic negativity directly in terms
of the covariance matrix. To this end, it is important to note
that on the level of covariance matrices the transposition is
reflected by time reversal which is a transformation that leaves
the positions invariant but reverses all momenta, $\hat q \mapsto \hat q$,
$\hat p \mapsto - \hat p$. The partial transposition is then the application
of this time-reversal transformation to a subsystem, i.e., one
party. Let us now consider a system made up of $m+n$ oscillators,
where $m$ oscillators are held by party $A$ and $n$ oscillators by
party $B$. Applying  time reversal to the $n$ oscillators held by
party $B$, the covariance matrix will be transformed to a real
symmetric matrix $\gamma^{T_B}$ given by
\begin{equation}
    \gamma^{T_B}= P\gamma  P,
\end{equation}
where
\begin{equation}\label{f}
    P= {\id}_{m+n}\oplus
    \left[\begin{array}{cc} \id_m & 0 \\ 0 & -\id_n \end{array}\right] .
\end{equation}
The $2n\times 2n$-matrix $\gamma^{T_B}$ is the matrix collecting
the second moments of the partial transpose $\rho^{T_B}$ of
$\rho$. The logarithmic negativity is then given by
\begin{equation}\label{logneg}
    N=-\sum_{j=1}^{m+n}\log_2(\min(1,|\gamma_j|)) \, ,
\end{equation}
where the $\gamma_j$ are the symplectic eigenvalues of
$\gamma^{T_B}$. For a general covariance matrix, $\gamma_j$ arises
in the diagonalization of $\gamma$ using symplectic
transformations, i.e., transformations $S$ that preserve the CCR
so that $S\sigma S^T=\sigma$. The resulting diagonal matrix is the
Williamson normal form of a covariance matrix whose diagonal
elements are the symplectic eigenvalues. It is sometimes useful to
know that the symplectic eigenvalues can be obtained directly
without explicit diagonalization of the matrix as the positive
square roots of the usual eigenvalues of $-\sigma \gamma \sigma
\gamma$ \cite{Vidal W 02}.

For all Hamiltonians that are quadratic in the canonical position
and momentum operators the ground state is an important example of
a Gaussian state. For a Hamilton operator of the form
\begin{equation}\label{hamiltonian}
    {\hat H} = \frac{1}{2} R^T \left[ \begin{array}{cc}
    V & 0 \\
    0 & T \end{array} \right] R
\end{equation}
we find that the covariance matrix of the ground state is given by
\begin{equation}
    \gamma = \sqrt{TV^{-1}}\oplus \sqrt{VT^{-1}}
\end{equation}
which reduces to
\begin{equation}
    \gamma = \sqrt{V^{-1}}\oplus \sqrt{V}
\end{equation}
when $T=\id_n$. If on the other hand, as for the interaction in
case (c), we have $T=V$, then the ground state is given by
$\gamma=\id_n\oplus \id_n$, which is the same as the ground-state
of $n$ non-interacting harmonic oscillators.

The primary aim of this work is the investigation of the dynamical
properties of the system of harmonic oscillators and the evolution
of entanglement properties under such dynamics. The dynamics of
the covariance matrix under a Hamiltonian quadratic in position
and momentum operators can be obtained straightforwardly from the
Heisenberg equation
\begin{equation}
    \frac{d}{dt} \hat X(t) = i [{\hat H},\hat X] .
\end{equation}
 For our time-independent Hamiltonian Eq.\
 (\ref{hamiltonian}), this leads to the covariance matrix at time $t$ as
\begin{eqnarray}
    \left[
\begin{array}{cc}
  \gamma_{XX}(t) & \gamma_{XP}(t)  \\
  \gamma_{PX}(t)  & \gamma_{PP}(t)  \\
\end{array}\right]
&=&
\exp\left({\left[
\begin{array}{cc}
  0 & T \\
  -V & 0 \\
\end{array}
\right] t}\right) \left[
\begin{array}{cc}
  \gamma_{XX} & \gamma_{XP} \\
  \gamma_{PX} & \gamma_{PP} \\
\end{array}
\right] \nonumber \\
&\times & \exp \left({\left[
\begin{array}{cc}
  0 & -V \\
  T & 0 \\
\end{array}
\right] t}\right).\label{num}
\end{eqnarray}
 Equipped with these tools we can now proceed to the analysis
of the entanglement dynamics of the harmonic chain.

\section{The equations of motion}
In two separate subsections, we provide the explicit solutions to
the equations of motion for the two coupling models (b) and (c)
that will be investigated both analytically and numerically in the
remainder of the paper.

\subsection{Harmonic oscillators coupled by springs}
This model is characterized by a Hamiltonian of the form
\begin{equation}\label{ham}
    \hat{H}_{\rm Spring}= \frac{1}{2}\sum_{k=1}^{M} {\hat q}_k^2 + {\hat p}_k^2 +
    c\left(\hat{q}_{k+1}-\hat{q}_k\right)^2 \, .
\end{equation}
Note that for the moment we neglect the decoupled additional
$0$-{th} oscillator. In the following we will provide an explicit
form for the equations of motion for the canonical positions and
momenta in the Heisenberg picture. To this end we can diagonalise
this Hamiltonian by introducing the normal coordinates
\begin{eqnarray}
    {\hat q}_n &=& \frac{1}{\sqrt{M}} \sum_{m=1}^{M} e^{\frac{2\pi i n
    m}{M}}{\hat Q}_m, \label{normalcoordq} \\ 
    {\hat p}_n &=& \frac{1}{\sqrt{M}} \sum_{m=1}^{M} e^{-\frac{2\pi i n m}{M}}
    {\hat P}_m \, .\nonumber \label{normalcoordp}
\end{eqnarray}
This leads to
\begin{equation}
    H_{\rm Spring} = \frac{1}{2} \sum_{s=1}^{M} \left( {\hat P}_s {\hat P}_s^{\dagger}
     + \omega_s^2 {\hat Q}_s{\hat Q}_s^{\dagger}\right)
\end{equation}
where
    $\omega_s^2 = 1+4c\sin^2(\pi s/M)$.
Here we have used the fact that ${\hat Q}_u = {\hat
Q}_{-u}^\dagger$,${\hat P}_u={\hat P}_{-u}^\dagger$. We introduce
the annihilation operators
\begin{eqnarray}
    {\hat a}_s &=& \frac{1}{\sqrt{2\omega_s}}(\omega_s {\hat Q}_s + i
    {\hat P}_s^{\dagger}), 
\end{eqnarray}
so that the Hamiltonian takes the form
\begin{equation}\label{ham1/2}
    H_{\rm Spring}=\sum_{s=1}^{M}\omega_s\Bigl
    (\hat a_s^\dagger \hat a_s+\frac{1}{2}\Bigr) \; .
\end{equation}
In the Heisenberg picture the annihilation and creation operators
evolve according to  ${\hat a}_s(t) = e^{-i\omega_s t} {\hat
a}_s(0)$ and ${\hat a}_s^{\dagger}(t) = e^{+i\omega_s t} {\hat
a}_s^{\dagger}(0)$.  Separating the real and imaginary parts,
we get
\begin{eqnarray}\label{QP}
    {\hat Q}_s(t) &=& {\hat Q}_s(0) \cos(\omega_s t) + \frac{1}{\omega_s}
    {\hat P}^{\dagger}_s(0) \sin( \omega_s t),\\
    {\hat P}_s(t) &=& -\omega_s {\hat Q}_s^{\dagger}(0) \sin (\omega_s t) +
    {\hat P}_s(0) \cos (\omega_s t )\; .\nonumber
\end{eqnarray}
Substituting these into Eq.\  (\ref{normalcoordq}), we obtain the
time evolution for the original position and momentum operators
\begin{eqnarray}
    {\hat q}_n(t) &=& \sum_{r=1}^{M} \left( {\hat q}_r(0) f_{r-n}(t) + {\hat p}_r(0) g_{r-n}(t)
    \right), \\
    {\hat p}_n(t) &=& \sum_{r=1}^{M} \big( {\hat q}_r(0) {\dot f}_{r-n}(t) + {\hat p}_r(0) f_{r-n}(t)
     \big),\nonumber
\end{eqnarray}
where we have defined the useful functions
\begin{eqnarray}
    f_k(t) &=& \frac{1}{M} \sum_{m=1}^{M} e^{\frac{2\pi i m
    k}{M}}\cos(\omega_m t),
    \label{useful1}\\
    g_k(t) &=& \frac{1}{M} \sum_{m=1}^{M} e^{\frac{2\pi i m
    k}{M}}\frac{\sin(\omega_m t)}{\omega_m}  \; .\nonumber
\end{eqnarray}
In the entire paper, dots will denote time derivatives. Defining
the covariance matrix elements to be (once we ignore the
displacements $\langle {\hat q}_i\rangle_\rho$)
\begin{equation}
    \gamma_{q_n q_m} = 2 \mbox{Re}\;
    \mbox{tr}
    [\rho{\hat q}_n{\hat q}_m],
\end{equation}
we find their values at time $t$ as
\begin{widetext}
\begin{eqnarray}
  \gamma_{q_n q_m}(t) &=& \sum_{r,s=1}^{M}
    \Bigl(
    f_{r-n}(t)f_{s-m}(t)\gamma_{q_r q_s} + g_{r-n}(t)g_{s-m}(t)\gamma_{p_r p_s}
+f_{r-n}(t)g_{s-m}(t)\gamma_{q_r p_s}
    + g_{r-n}(t)f_{s-m}(t)\gamma_{p_r q_s} \Bigr), \\
    \gamma_{q_n p_m}(t) &=& \sum_{r,s=1}^{M}\Bigl(
    f_{r-n}(t){\dot f}_{s-m}(t)\gamma_{q_r q_s} + g_{r-n}(t)f_{s-m}(t)
    \gamma_{p_r p_s} 
+ f_{r-n}(t)f_{s-m}(t)\gamma_{q_r p_s}
    + g_{r-n}(t){\dot f}_{s-m}(t)\gamma_{p_r q_s} \Bigr),\nonumber \\
   \gamma_{p_n p_m}(t) &=&  \sum_{r,s=1}^{M} \Bigl(
    {\dot f}_{r-n}(t){\dot f}_{s-m}(t)\gamma_{q_r q_s} + f_{r-n}(t)f_{s-m}(t)\gamma_{p_r p_s} 
 + {\dot f}_{r-n}(t)f_{s-m}(t)\gamma_{q_r p_s}
    + f_{r-n}(t){\dot f}_{s-m}(t)\gamma_{p_r q_s}\Bigr),\nonumber
\end{eqnarray}
\end{widetext}
where the $\gamma$ on the right hand side are the initial values
of the covariance matrix elements.

\subsection{Hamiltonian in the rotating wave approximation}
This model is characterized by the Hamiltonian
\begin{equation}
    \hat{H}_{\rm RWA} = \frac{1}{2} \sum_{k=1}^{M}
    {\hat q}_k^2 + {\hat p}_k^2 + \frac{c}{2} \left(\hat{q}_{k+1}
    -\hat{q}_k\right)^2 + \frac{c}{2} \left(\hat{p}_{k+1}-\hat{p}_k\right)^2 \, ,
\end{equation}
and in the following, we can carry out the analysis along the same
lines as in the previous subsection. Again, employing the normal
coordinates Eqs.\ (\ref{normalcoordq}) we obtain
\begin{equation}
    {\hat H}_{\rm RWA} = \frac{1}{2} \sum_{s=1}^{M} \Omega_s^2\left( {\hat P}_s {\hat P}_s^{\dagger}
    + {\hat Q}_s{\hat Q}_s^{\dagger}\right)
\end{equation}
where we now have
\begin{equation}
    \Omega_s^2 =  1+2c\sin^2\biggl(\frac{\pi s}{M}\biggr) \, .
\end{equation}
Introducing the annihilation and creation operators
\begin{eqnarray}
    {\hat a}_s &=& \frac{1}{\sqrt{2}}({\hat Q}_s + i
    {\hat P}_s^{\dagger}), \;\;\;\;\;
    {\hat a}_s^{\dagger} = \frac{1}{\sqrt{2}}({\hat Q}_s^{\dagger} - i
    {\hat P}_s)
\end{eqnarray}
the Hamiltonian assumes a form
\begin{equation}
    {\hat H}_{\rm RWA} = \sum_{s=1}^M\Omega_s^2\left({\hat a}_s^\dag {\hat a}_s+\frac{1}{2}
    \right).
\end{equation}
In the Heisenberg picture the annihilation and creation operator
then evolve in time as ${\hat a}_s(t) = e^{-i\Omega^2_s t} {\hat
a}_s(0)$ and ${\hat a}_s^{\dagger}(t) = e^{+i\Omega^2_s t} {\hat
a}_s^{\dagger}(0)$. Separating again real and imaginary parts we
obtain
\begin{eqnarray}
    {\hat Q}_s(t) &=& {\hat Q}_s(0) \cos(\Omega_s^2 t) + {\hat P}^{\dagger}_s(0) \sin (\Omega_s^2 t),\\
    {\hat P}_s(t) &=& -{\hat Q}_s^{\dagger}(0) \sin(\Omega_s^2 t) + {\hat P}_s(0) \cos (\Omega_s^2
    t)\, .\nonumber
  \end{eqnarray}
Transforming back to the original position and momentum operators
we find
\begin{eqnarray}
    {\hat q}_n(t) &=& \sum_{r=1}^{M} \left({\hat q}_r(0) F_{r-n}(t) + {\hat p}_r(0) G_{r-n}(t)
    \right),\\
    {\hat p}_n(t) &=& \sum_{r=1}^{M} \left(-{\hat q}_r(0) G_{r-n}(t) + {\hat p}_r(0) F_{r-n}(t) \right),\nonumber
\end{eqnarray}
\begin{widetext}
where we have defined the functions $F_k$ and $G_k$ as
\begin{eqnarray}
     F_k(t) &=& \frac{1}{M} \sum_{m=1}^{M} e^{\frac{2\pi i m
    k}{M}}\cos(\Omega_m^2 t), \,\,\,\,
    G_k(t) = \frac{1}{M} \sum_{m=1}^{M} e^{\frac{2\pi i m
    k}{M}}\sin(\Omega_m^2 t) \, .
\end{eqnarray}
Note that these functions are slightly simpler than the
corresponding ones in Eqs.\ (\ref{useful1}) as they lack the
frequency $\Omega_s$ in the denominator. The covariance matrix
elements vary in time as
\begin{eqnarray}
  \gamma_{q_n q_m}(t) &=& \sum_{r,s=1}^{M}
    \Bigl(
    F_{r-n}(t)F_{s-m}(t)\gamma_{q_r q_s} + G_{r-n}(t)G_{s-m}(t)\gamma_{p_r p_s}
+ F_{r-n}(t)G_{s-m}(t)\gamma_{q_r p_s}
    + G_{r-n}(t)F_{s-m}(t)\gamma_{p_r q_s} \Bigr),\nonumber\\
    \gamma_{q_n p_m}(t) &=& \sum_{r,s=1}^{M}\Bigl(
    -F_{r-n}(t)G_{s-m}(t)\gamma_{q_r q_s} + G_{r-n}(t)F_{s-m}(t)
    \gamma_{p_r p_s} 
   +F_{r-n}(t)F_{s-m}(t)\gamma_{q_r p_s}
    - G_{r-n}(t)G_{s-m}(t)\gamma_{p_r q_s} \Bigr),\nonumber \\
   \gamma_{p_n p_m}(t) &=&  \sum_{r,s=1}^{M} \Bigl(
    G_{r-n}(t)G_{s-m}(t)\gamma_{q_r q_s} + F_{r-n}(t)F_{s-m}(t)\gamma_{p_r p_s}  
  - G_{r-n}(t)F_{s-m}(t)\gamma_{q_r p_s}
    - F_{r-n}(t)G_{s-m}(t)\gamma_{p_r q_s}\Bigr), \nonumber\\
\end{eqnarray}

\section{Propagation of entanglement along the chain}
We would like to investigate the capacity of the harmonic chain
for transmission of quantum information. The clearest signature
for the ability to transmit quantum information and coherence is
the proof of the ability to transmit one constituent part of an
entangled pair of oscillators through the chain. To analyze the
situation we require the equations of motion for the covariance
matrix. Let us assume the existence of a distinguished oscillator
$0$ which is entirely decoupled from the others. We imagine that
at time $t=0$ this oscillator is prepared in a two-mode squeezed
state with the first oscillator of the chain while all other
oscillators are assumed to be in their respective ground state
(assuming no interaction). The initial conditions then read
\begin{eqnarray}
    \gamma_{q_0 q_0} &=& \gamma_{q_1 q_1} = \gamma_{p_0 p_0}
    = \gamma_{p_1 p_1} = \cosh(r),\label{ini1} \,\,\,\,\,
    \gamma_{q_0 q_1} = - \gamma_{p_0 p_1} = \sinh(r), \\
    \gamma_{q_s q_s} &=& \gamma_{p_s p_s} = 1 \;\; \mbox{for all}\,\;  s>1,\nonumber\\
    \gamma_{q_r p_s} &=& 0 \; .\nonumber
\end{eqnarray}
The $0$-{th} oscillator for the interaction via springs will obey
a free time evolution which is given by (using Eqs.\ (\ref{QP}) and
noting that $\hat q_0=\hat Q_0$ and $\hat p_0=\hat P_0$)
\begin{eqnarray}
    \hat q_0(t) &=& \hat q_0(0)\cos(\omega_0 t)+\frac{\hat p_0(0)\sin(\omega_0
    t)}{\omega_0}, \,\,\,
    \hat p_0(t) = -\omega_0 \hat q_0(0)\sin(\omega_0 t)+\hat p_0(0)\cos(\omega_0
    t) \, .
\end{eqnarray}
Similarly, for the RWA interaction
\begin{eqnarray}
    \hat q_0(t) &=& \hat q_0(0)\cos(\Omega^2_0 t)+\hat p_0(0)\sin(\Omega^2_0t),\,\,\,
    \hat p_0(t) = -\hat q_0(0)\sin(\Omega^2_0 t)+\hat p_0(0)\cos(\Omega^2_0
    t) \, .
\end{eqnarray}
Note, however, that they correspond to local unitary rotations on
the $0$-{th} oscillator which do not affect the entanglement
between this oscillator and the remaining ones. To simplify the
expressions we will therefore omit this time evolution in the
following. Again we will treat the two types of interactions
described by ${\hat H}_{\rm Spring}$ and ${\hat H}_{\rm RWA}$
separately. The next two subsections establish the analyutical
expressions for the time evolution of the entanglement between the
$0$-th oscillator and the $n$-th oscillator.

\subsection{Harmonic oscillators coupled by springs}
In this case, employing the special form of the initial conditions
for the system Eqs.\ (\ref{ini1}), the elements of the covariance
matrix describing the $0$-th and the $n$-th oscillator at time
$t$ are given by
\begin{eqnarray}
   &\gamma_{q_0 q_0}(t) = \gamma_{p_0 p_0}(t) = \cosh(r),\,\,\,\,\,\,
   &\gamma_{q_0 q_n}(t) =\sinh(r) f_{n-1}(t), \nonumber\\
& \gamma_{p_0 p_n}(t)= - \sinh(r)
f_{n-1}(t),\nonumber
  & \gamma_{q_0 p_0}(t) = 0,\\
   & \gamma_{q_0 p_n}(t) = \sinh(r) {\dot f}_{n-1}(t),
 &\gamma_{q_n p_0}(t) = -\sinh(r) g_{n-1}(t),\nonumber\\
 \end{eqnarray}
 and
 \begin{eqnarray}
 \gamma_{q_n q_n}(t) &=&
(\cosh(r)-1)(f^2_{n-1}(t)+g^2_{n-1}(t))
    + \sum_{s=1}^{M} (f^2_{n-s}(t) + g^2_{n-s}(t)),\\
    \gamma_{q_n p_n}(t) &=& (\cosh(r)
    -1)(f_{n-1}(t){\dot f}_{n-1}(t)+g_{n-1}(t)f_{n-1}(t))
%
+ \sum_{s=1}^{M}f_{n-s}(t){\dot f}_{n-s}(t) +
g_{n-s}(t)f_{n-s}(t),\nonumber\\
    \gamma_{p_n p_n}(t) &=& (\cosh(r)-1)({\dot
f}^2_{n-1}(t)+f^2_{n-1}(t))
    + \sum_{s=1}^{M} ({\dot f}^2_{n-s}(t) + f^2_{n-s}(t)) \, .\nonumber
\end{eqnarray}
In the limit of a chain of infinite length, i.e., when
$M\rightarrow\infty$, we find
\begin{eqnarray}
     &\gamma_{q_0 q_0}(t) = \gamma_{p_0 p_0}(t) = \cosh(r), \,\,\,\,\,\,
     &\gamma_{q_0 q_n}(t) = -\gamma_{p_0 p_n}(t)
=\sinh(r)f_{n-1}(t),  \\
    & \gamma_{q_0 p_0}(t) = 0,
     &\gamma_{q_0 p_n}(t) = \sinh(r) {\dot f}_{n-1}(t), \nonumber\\
    &  \gamma_{q_0 p_n}(t) = \sinh(r) {\dot f}_{n-1}(t),
    & \gamma_{q_n p_0}(t) = -\sinh(r)g_{n-1}(t), \nonumber
\end{eqnarray}
and
\begin{eqnarray}
     \gamma_{q_n q_n}(t) &=& (\cosh(r)-1)
(f^2_{n-1}(t)+g^2_{n-1}(t)) + a_{nn}(t) + d_{nn}(t),  \\
     \gamma_{q_n p_n}(t) &=& (\cosh(r)
    -1)(f_{n-1}(t){\dot f}_{n-1}(t)+g_{n-1}(t)f_{n-1}(t)) + b_{nn}(t) + e_{nn}(t), \nonumber\\
     \gamma_{p_n p_n}(t) &=& (\cosh(r)-1)({\dot
f}^2_{n-1}(t)+f^2_{n-1}(t)) + c_{nn}(t) + a_{nn}(t). \nonumber
\end{eqnarray}
Here we have employed the definitions $\zeta= {c}/{(1+2c)}$ and
$\Omega=\sqrt{1+2c}$, and introduced the functions
\begin{eqnarray}
    f_k(t) &=& \frac{1}{\pi} \int_{0}^{\pi} d\phi \cos(k\phi)\,
    \cos \left(\Omega t \sqrt{1-2\zeta\cos\phi}\right),  \,\,\,
    g_k(t) = \frac{1}{\pi} \int_{0}^{\pi} d\phi \cos(k\phi)\,
    \frac{\sin \left(\Omega t \sqrt{1-2\zeta\cos\phi}\right)}{\Omega \sqrt{1-2\zeta\cos\phi}}
    ,
    \nonumber
\end{eqnarray}
and
\begin{eqnarray}
    a_{nn}(t) &=& \frac{1}{2}\left(1 + \frac{1}{\pi}
    \int_{0}^{\pi} d\phi \cos\left(2\Omega t \sqrt{1-2\zeta\cos\phi}\right)\right),     \\[0.2cm]
    b_{nn}(t) &=& -\frac{1}{2\pi}\int_{0}^{\pi} d\phi\; \Omega\sqrt{1-2\zeta\cos\phi}
    \sin\left(2\Omega t \sqrt{1-2\zeta\cos\phi}\right), \nonumber\\[0.2cm]
    c_{nn}(t) &=& \frac{\Omega^2}{2}-\frac{\Omega^2}{2\pi}\int_{0}^{\pi} d\phi
    (1-2\zeta\cos\phi)\cos \left(2\Omega t\sqrt{1-2\zeta\cos\phi}\right), \nonumber\\[0.2cm]
    d_{nn}(t) &=& \frac{1}{2} \frac{1}{\sqrt{1+4c}}-
    \frac{1}{2\pi}\int_{0}^{\pi} d\phi \frac{\cos \left(2\Omega t
    \sqrt{1-2\zeta\cos\phi}\right)}{\Omega^2(1-2\zeta\cos\phi)}, \nonumber\\[0.2cm]
    e_{nn}(t) &=& \frac{1}{2\pi}\int_{0}^{\pi} d\phi \frac{\sin \left(2\Omega t
    \sqrt{1-2\zeta\cos\phi}\right)}{\Omega\sqrt{1-2\zeta\cos\phi}} \, .\nonumber
\end{eqnarray}
While the above set of equations determines the time evolution
exactly, they do not provide very much direct insight into the
dynamics of the system. In the following we will show however,
that we can obtain very good and compact approximations to the
above exact solution in terms of elementary functions. While the
following derivation is not rigorous in the sense that it does not
provide error bounds, a numerical comparison between the
approximations and the exact results shows the impressive quality
of the approximate solution.
As a first step, we expand the functions $f_k$ and $g_k$ to first
order in $\zeta$,
\begin{eqnarray}
    f_k(t) &\cong& \frac{1}{\pi} \int_{0}^{\pi} d\phi \cos (k\phi)\, \cos \left(\Omega t (1-\zeta\cos\phi)\right), \,\,\,
    g_k(t) \cong \frac{1}{\pi} \int_{0}^{\pi} d\phi \cos (k\phi)\,
    \sin \left(\Omega t (1-\zeta\cos\phi)\right), 
\end{eqnarray}
and also drop a factor of ${1}/{\Omega \sqrt{1-2\zeta\cos\phi}}$
in $g_k$. In the following we will employ Bessel functions which
satisfy the relations
\begin{eqnarray}
    \cos(x\cos s) &=& J_0(x) + 2 \sum_{n=1}^{\infty} J_{2n}(x)
    \cos(2ns)\cos(n\pi),\nonumber \\
    \sin(x\cos s) &=& 2 \sum_{n=0}^{\infty} J_{2n+1}(x)
    \cos((2n+1)s)\cos(n\pi) \, .
\end{eqnarray}
On using trigonometrical addition theorems one finds that in first
order
\begin{eqnarray}
    f_k(t) &\cong& J_k(\zeta\Omega t) \cos\Big( \Omega t - \frac{\pi k}{2}
    \Big), \;\;\;\;\; 
    g_k(t)  \cong   J_k(\zeta\Omega t) \sin\Big(\Omega t - \frac{\pi
    k}{2}\Big)\, .\label{approxg}
\end{eqnarray}
A further crucial approximation replaces the time-dependent
quantities $a_{nn}(t)$, $b_{nn}(t)$, $c_{nn}(t)$, $d_{nn}(t)$, $e_{nn}(t)$ by
their time averages, i.e.,
\begin{eqnarray}
    a_{nn}(t) \mapsto
    \lim_{T\rightarrow \infty}\frac{1}{T}\int_{0}^T dt a_{nn}(t) &=&
    \frac{1}{2}, \,\,\,\,
    b_{nn}(t) \mapsto
     \lim_{T\rightarrow \infty}\frac{1}{T}\int_{0}^T dt b_{nn}(t) = 0,
     \\
    c_{nn}(t) \mapsto
       \lim_{T\rightarrow \infty}\frac{1}{T}\int_{0}^T dt c_{nn}(t) &=& \frac{\Omega^2}{2}, \nonumber
       \,\,\,\,
    d_{nn}(t) \mapsto
     \lim_{T\rightarrow \infty}\frac{1}{T}\int_{0}^T dt d_{nn}(t) = \frac{1}{2}\frac{1}{\sqrt{1+4c}}, \nonumber\\[0.1cm]
    e_{nn}(t) \mapsto
      \lim_{T\rightarrow \infty}\frac{1}{T}\int_{0}^T dt e_{nn}(t) &=& 0. \nonumber
\end{eqnarray}
With all these approximations we finally obtain
\begin{eqnarray}
    \gamma_{q_0 q_0}(t) &\cong& \gamma_{p_0 p_0}(t) \cong \cosh(r),\\
    \gamma_{q_0 q_n}(t) &\cong& \gamma_{p_0 p_n}(t)      \cong \sinh(r)
    J_{n-1}(\zeta\Omega t) \cos\Big( \Omega t - \frac{\pi (n-1)}{2} \Big),\nonumber\\
    \gamma_{q_0 p_0}(t) &\cong& \gamma_{q_n p_n}(t) \cong 0,\nonumber\\
    \gamma_{q_0 p_n}(t) &\cong& \gamma_{q_n p_0}(t) 
     \cong  -\sinh (r)
    J_{n-1}(\zeta\Omega t) \sin\Big( \Omega t - \frac{\pi (n-1)}{2} \Big),\nonumber\\
    \gamma_{q_n q_n}(t) &\cong& (\cosh(r) -1)J_{n-1}^2(\zeta\Omega t)
     + \frac{1}{2} + \frac{1}{2}\frac{1}{\sqrt{1+4c}},\nonumber\\
    \gamma_{p_n p_n}(t) &\cong& (\cosh(r)
    -1)J_{n-1}^2(\zeta\Omega t) + 1+c. \nonumber
\end{eqnarray}
Numerical comparisons in later sections will show that these
approximate solutions are very good approximations when $r$ is not
too small.
We have so far collected all the elements of the covariance matrix
involving the $0$-th and the $n$-th oscillator. Since we will
investigate the entanglement properties between the two
oscillators, we can trace out the rest of the oscillators which
leaves us with the covariance matrix of the reduced system
comprising only two oscillators. Employing the ordering
$(\hat q_0,\hat p_0,\hat q_n,\hat p_n)$ we find
\begin{equation}
\gamma_{\rm Spring} \cong \left[
\begin{array}{cc}
    A & D \\
    D^T & B
\end{array}
\right],
\end{equation}
where
\begin{eqnarray}
  A &=&\left[
\begin{array}{cccc}
  \cosh(r) &  0 \\
  0    &   \cosh(r) \\
  \end{array}
\right],
\\[0.2cm]
  B &=&\left[
\begin{array}{cccc}
  (\cosh(r)-1)J_s(t)^2+1/2+(1+4c)^{-1/2}/2 & 0\\
   0 & (\cosh(r)-1)J_s(t)^2+1+c
  \end{array}
\right], \nonumber \\[0.1cm]
  D &=&\left[
\begin{array}{cccc}
\sinh(r)J_s(t)\cos\Phi_s(t) & -\sinh(r)J_s(t)\sin\Phi_s(t) \\
-\sinh(r)J_s(t)\sin\Phi_s(t) & -\sinh(r)J_s(t)\cos\Phi_s(t)
  \end{array}
\right], \nonumber
\end{eqnarray}
where we have used the abbreviations $\Phi_s(t)=\Omega t-\pi
(n-1)/2$ and $J_s(t)=J_{n-1}(\zeta\Omega t)$. From this explicit
form for the covariance matrix of the $0$-th and $n$-th
oscillator we can now determine the symplectic eigenvalues as
solutions of the polynomial \cite{Vidal W 02}
\begin{equation}\label{quadratic}
    \eta^4-(\det(A)+\det(B)-2\det(D))\eta^2+\det(\gamma_{\rm Spring})=0.
\end{equation}
The solution is given by
\begin{equation}\label{etaspring}
    \eta^2_{\rm Spring} = \frac{1}{4}\left( y_1 - y_2\right),
\end{equation}
where we have
\begin{eqnarray}
  y_1 &=& 2+c+(1+c)(1+4c)^{-\frac{1}{2}} -
(5+2c+(1+4c)^{-\frac{1}{2}})J_s^2 + 3 J_s^4  \nonumber\\
  && +J_s^2(3+2c+(1+4c)^{-\frac{1}{2}} - 4J_s^2)\cosh (r)
+(1+J_s^2)^2\cosh (2r) , \\
     y_2 &=& \sqrt{w+y_1^2}, \nonumber
\end{eqnarray}
and
\begin{equation}
     w =
-8[2J_s^2+(1+(1+4c)^{-\frac{1}{2}}-2J_s^2)\cosh(r)][J_s^2+(1+c-J_s^2)\cosh(r)].
\end{equation}
Note that we have dropped the time argument in $J_s(t)$ to make
the expressions appear more compact. The logarithmic negativity,
finally, is in this approximation given by
%
%
%
%
\begin{equation}
    N_{\rm Spring}(t)\cong-\log_2(\min(|\eta_{\rm Spring}|,1)).
\end{equation}
Note that the other symplectic eigenvalue is greater than one.
\subsection{Hamiltonian in rotating wave approximation}
For the Hamiltonian $H_{\rm RWA}$ one proceeds along very similar
lines, i.e., taking the limit $M\rightarrow \infty$ to find
\begin{eqnarray}
    F_k(t) &=& \frac{1}{\pi} \int_{0}^{\pi} d\phi \cos (k\phi)
    \cos[\Omega_{\rm RWA}^2 t (1-\zeta_{\rm RWA} \cos\phi)] ,\\
    G_k(t) &=& \frac{1}{\pi} \int_{0}^{\pi} d\phi \cos (k\phi)
    \sin[\Omega_{\rm RWA}^2 t (1-\zeta_{\rm RWA} \cos\phi)],\nonumber
\end{eqnarray}
with $\Omega_{\rm RWA}=\sqrt{1+c}$ and $\zeta_{\rm RWA}=c/(1+c)$,
which are exactly
\begin{eqnarray}
   F_n(t) &=& J_n(ct)\cos(\Omega_{\rm RWA}^2t-\pi n/2), \,\,\,\,
     G_n(t)  = J_n(ct)\sin(\Omega_{\rm RWA}^2t-\pi n/2) \, .
\end{eqnarray}
we find the covariance matrix elements to be
\begin{eqnarray}
  \gamma_{q_0q_0}(t) &=& \gamma_{p_0p_0}(t) = \cosh(r) , \\
  \gamma_{q_0p_0}(t) &=& \gamma_{q_np_n}(t) = 0 , \nonumber\\
  \gamma_{q_0q_n}(t) &=& -\gamma_{p_0p_n}(t) = \sinh (r) F_{n-1}(t) , \nonumber\\
  \gamma_{q_0p_n}(t) &=& \gamma_{q_np_0}(t) = -\sinh (r) G_{n-1}(t), \nonumber \\
  \gamma_{q_nq_n}(t) &=& \gamma_{p_np_n}(t) = (\cosh(r) -1)(F_{n-1}^2(t)+G_{n-1}^2(t))+1.  \nonumber
\end{eqnarray}
Note that now the terms corresponding to
$a_{nn}(t),b_{nn}(t),c_{nn}(t),d_{nn}(t)$, and $e_{nn}(t)$ do not
need to be approximated as all time-dependencies cancel each other
off conveniently in the expressions  as opposed to the spring
case. Again we can write the covariance matrix of the reduced
system comprising only the $0$-th and the $n$-th oscillator
employing the ordering $(\hat q_0,\hat p_0,\hat q_n,\hat p_n)$, and find
\begin{equation}
\gamma_{\rm RWA} = \left[
\begin{array}{cc}
    A & D \\
    D^T & B
\end{array}
\right]
\end{equation}
with
\begin{eqnarray}
  A &=&\left[
\begin{array}{cccc}
  \cosh(r) &  0 \\
  0    &   \cosh(r) \\
  \end{array}
\right], \\[0.1cm]
  B &=&\left[
\begin{array}{cccc}
  (\cosh(r)-1)J_{\rm RWA}(t)^2 + 1 & 0\\
   0 & (\cosh(r)-1)J_{\rm RWA}(t)^2 + 1
  \end{array}
\right], \nonumber\\[0.1cm]
  D &=&\left[
\begin{array}{cccc}
    \sinh(r)J_{\rm RWA}(t)\cos\Phi_{\rm RWA}(t) &
    -\sinh(r)J_{\rm RWA}(t)\sin\Phi_{\rm RWA}(t)\\
    -\sinh(r)J_{\rm RWA}(t)\sin\Phi_{\rm RWA}(t) &
    -\sinh(r)J_{\rm RWA}(t)\cos\Phi_{\rm RWA}(t)
  \end{array}
\right], \nonumber
\end{eqnarray}
\end{widetext}
where we have used the abbreviations $\Phi_{\rm
RWA}(t)=\Omega_{\rm RWA}^2 t-\pi (n-1)/2$ and $J_{\rm RWA}(t) =
J_{n-1}(ct)$. Note that these expressions are very similar to
those for the first interaction type. The symplectic eigenvalues
are given by the solutions of
\begin{equation}
    \eta_{\rm RWA}^4-(\det(A)+\det(B)-2\det(D))\eta_{\rm RWA}^2+\det(\gamma_{\rm RWA})=0.
\end{equation}
This gives rise to
\begin{equation}
    \eta^2_{\rm RWA} = \frac{1}{2}\left( z_1 - z_2\right),
\end{equation}
with
\begin{eqnarray}
    z_1 &=& (1+J_{\rm RWA}^2)\cosh^2(r)+2J_{\rm RWA}^2\sinh^2(r)\nonumber\\
&&+2J_{\rm RWA}^2(1-J_{\rm RWA}^2)\cosh(r)
  +(1-J_{\rm RWA}^2)^2 ,\nonumber \\
    z_2 &=& \sqrt{v+z_1^2}, 
\end{eqnarray}
where
\begin{equation}
    v=-4\left( (1-J_{\rm RWA}^2)\cosh(r)+J_{\rm RWA}^2 \right)^2.
\end{equation}
 Again we have dropped the explicit time dependence in
$J_{\rm RWA}$ for brevity of notation. The logarithmic negativity
is then given by
\begin{equation}
    N_{\rm RWA}(t)=-\log_2(\min(|\eta_{\rm RWA}|,1)) \, .
\end{equation}
Having prepared all the analytical work we need, we can now
proceed to investigate the entanglement dynamics of the harmonic
chain.

\section{Study of the entanglement dynamics of the harmonic chain}

This section studies numerically and analytically, various aspects
of the entanglement dynamics of the harmonic chain with various
initial states and geometrical arrangements for both types of
interactions. We begin by briefly revisiting the effect of the
spontaneous creation of entanglement which is obtained when the
interaction strength between the oscillators is globally changed
suddenly \cite{Eisert PBH 03}. This effect can only be observed in
a model in which the oscillators are coupled via springs. In the
rotating wave approximation this effect does not occur as both the
coupled and uncoupled chains have an identical ground state. Then
we move on to consider the propagation of entanglement through a
harmonic chain. Surprisingly, for harmonic oscillators coupled by
springs corresponding to a phonon model, we observe a
non-monotonic transfer efficiency in the initially prepared amount
of entanglement, i.e., an intermediate amount of entanglement is
transferred with the highest efficiency. In the rotating wave
approximation, the transfer efficiency is monotonic though equally
surprising. We provide analytical expressions for the propagation
speed of the entanglement through the chain together with
approximate analytical expressions for the transfer efficiency. We
also study the influence of imperfections such as finite
temperatures and varying coupling constants. Finally we study
different geometrical configurations that are analogous to quantum
optical devices such as beamsplitters and interferometers and
observe characteristic differences when initially thermal or
squeezed states are entering these devices. We propose ways in
which these devices may be in and out of action thereby allowing
for the creation of pre-fabricated quantum networks that can be
programmed by external switches.

\subsection{Spontaneous creation of entanglement}
\begin{figure}[h]
\begin{center}
\rotatebox{0}{\resizebox{!}{6cm}{\includegraphics{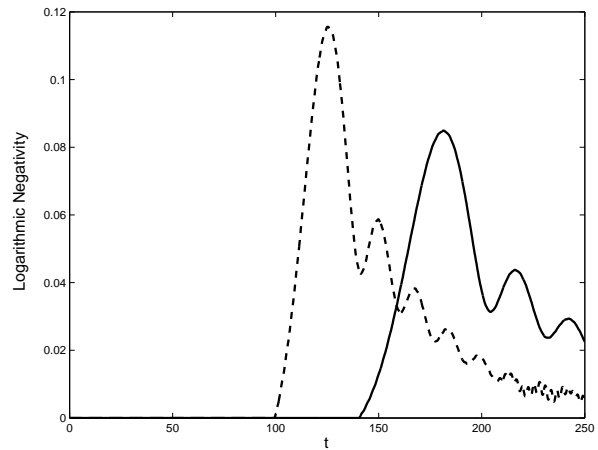}}}
\caption{\label{gs} The time evolution of
entanglement between the first and the last oscillator in an open
chain of length $30$ when all the oscillators are initially in the
ground state. The coupling strength $c=0.1$ has been chosen for
the solid line ($c=0.15$ for the dashed line).}
\end{center}
\end{figure}

We consider all the harmonic oscillators to be in the ground state
and initially uncoupled, i.e.,
\begin{equation}
    \gamma_{q_n  q_m}= \delta_{nm},\,\,\,\,
    \gamma_{q_n p_m}=0, \,\,\,\,
    \gamma_{p_n p_m}=\delta_{nm}.
\end{equation}
We suddenly switch on the interaction at time $t=0$ and observe
the dynamical evolution of entanglement. We do not observe any
entanglement with the $H_{\rm RWA}$ interaction (because we have
not included a mechanism to produce a spontaneous excitation, that is,
we do not have terms such as $a_k^\dag a_{k+1}^\dag$) so we will
exclusively deal with $\hat H_{\rm Spring}$. We are limiting our scope
to numerical results because the analytical expressions, while
they can be provided, are too complicated to yield any insight.
Note also that the approximations used below in the squeezed state
case cannot be applied here. A typical example of the time
evolution of entanglement is shown in Figure \ref{gs}. In an open
chain of length $30$ we study the time evolution of the
entanglement between the first and the last oscillator.
We observe no entanglement for a finite time until at a time $t_0$
one first encounters a build-up of entangled between the two
oscillators. This time $t_0$ is approximately given by
\begin{equation}
    t_0 \cong \frac{n}{2 \Omega \zeta},
\end{equation}
ie the time $t_0$ is approximately linear in $n$, the separation
of the oscillators. It should be noted that $t_0$ is half as large
as the time that is required for a perturbation at the first
oscillator to travel to the $n$-th oscillator. This suggests
that the origin of the entanglement between the $1^{st}$ and the
$n$-th oscillator arises from the interaction of those
oscillators exactly half-way in between. Their entanglement is
generated by the initial sudden switching on of the interaction
and then propagates through the chain. This idea will be further
corroborated in the later subsections when the propagation of
pre-prepared entanglement is considered. Furthermore one finds
that the dependence of the maximal degree of entanglement as
measured by the logarithmic negativity is approximately
proportional to $n^{-1/3}$ for large $n$ until reaches values of
about $10^{-2}$ when it begins to drop quite rapidly to vanish
entirely.  It should be noted that for any parameters of
the model there exists a finite $n$ such that the state of the
first and the $n$-th oscillator is separable for all times. For
the coupling value of $c=0.1$ this will happen for $n\cong 20000$.
Therefore, the value of largest separation is very large indeed
for reasonable coupling strengths. To appreciate that this is a
somewhat surprising behaviour, it should be contrasted with the
entanglement structure in the ground state of the chain. Then it
can be shown that for any chosen coupling strength two
distinguished oscillators are never in an entangled state, unless
they are immediately neighbouring \cite{Audenaert EPW 02}.

Here we have studied the special case of the entanglement between
the endpoint of an open chain. It should be noted that this is a
particularly favourable configuration. For a given distance between
oscillators one always obtains the largest amount of entanglement
when one places them at the opposite ends of an open chain. Two
oscillators in a very long chain with the same distance and with
positions well away from the ends of the chain will lead to
considerably smaller amounts of entanglement. Indeed, the amount
of entanglement will differ by approximately a factor of $4$. This
discrepancy is due to the fact that at the ends of the chain the
oscillators possess fewer neighbours with which they can become
entangled. As we discard all oscillators other than two any
entanglement with other oscillators will deteriorate the
entanglement between the distinguished oscillators. While this
does not explain the factor of $4$ quantitatively, it gives an
intuitive picture for the decrease of entanglement that will
discussed in more detail later on.

\subsection{Thermal state case}
\begin{figure}[h]
\begin{center}
\rotatebox{0}{\resizebox{!}{6cm}{\includegraphics{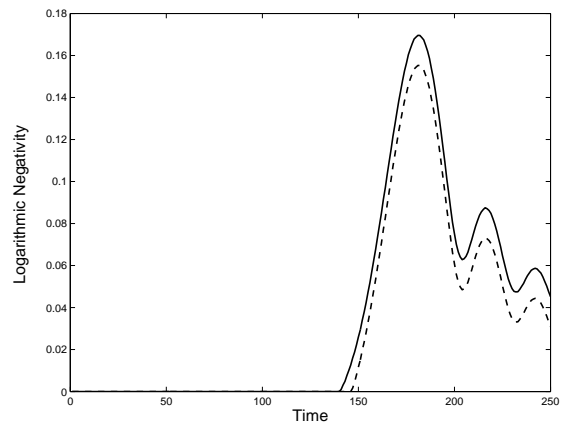}}}
\caption{\label{ts} The time evolution of
entanglement between the first and the last oscillator (with open
boundary conditions). We have fixed $c=0.1$, the chain consists of
$30$ oscillators, and $x= \omega/T$ for $x\geq 10$ (solid line)
and $x=6$ (dashed line).}
\end{center}
\end{figure}

In the previous section we have studied the entanglement dynamics
in an environment that is at zero temperature. This is reflected
by the fact that the initial state of our harmonic oscillators is
assumed to be the vacuum. We will now move one step further
towards a more realistic description by setting the initial state
as a thermal equilibrium state. As with the ground state case, we
do not establish entanglement for the $H_{\rm RWA}$ interaction as
a thermal state can be represented as a mixture of displaced
vacuum states which do not lead to any spontaneous entanglement in
the rotating wave approximation. Therefore we shall again only
consider the $H_{\rm Spring}$ interaction. The thermal equilibrium
state is given by
\begin{eqnarray}
    \gamma_{q_nq_m} &=& \delta_{nm}
    \Bigl(1+\frac{2}{e^{ \omega/T}-1}\Bigr), \\
    \gamma_{q_np_m} &=& 0, \nonumber\\
    \gamma_{p_np_m} &=&
    \delta_{nm}\Bigl(1+\frac{2}{e^{ \omega/T}-1}\Bigr). \nonumber
\end{eqnarray}
The equations of motion are then
\begin{eqnarray}
    \gamma_{q_n q_m}(t) &=& \Bigl(1+\frac{2}{e^{ \omega/T}-1}\Bigr) (a_{nm}(t)+d_{nm}(t)),\\
    \gamma_{q_n p_m}(t) &=& \Bigl(1+\frac{2}{e^{ \omega/T}-1}\Bigr) (b_{nm}(t)+e_{nm}(t)),\nonumber \\
    \gamma_{p_n p_m}(t) &=&\Bigl (1+\frac{2}{e^{ \omega/T}-1}\Bigr) (c_{nm}(t)+a_{nm}(t)).\nonumber
\end{eqnarray}
Appropriate values for the temperatures have to be obtained from
experiment in the particular context under consideration, but it
appears possible nowadays to achieve ratios $T/\omega\ll 1$ in
different physical systems (note that we have taken $\hbar=1$ and
$k=1$) such as nano-mechanical oscillators. We consider again an
open chain instead of a closed ring. This renders the analytical
treatment more difficult but makes no difference for the numerics.
The entanglement depends little on the temperature as long as the
mean thermal photon number is well below unity. Figure \ref{ts}
shows the temperature dependence of entanglement evolution.
We observe that down to temperatures corresponding to $x=10$, the
entanglement evolution is almost exactly the same as in the ground
state case. Only when $x<10$ do we see an effect of a finite
temperature. Even for $x=6$, we still have a significant portion
of entanglement albeit a small delay in the arrival of
entanglement. A more realistic scenario is dealt with in Ref.\
\cite{Eisert PBH 03}.

\begin{figure}[h]
\begin{center}
\rotatebox{0}{\resizebox{!}{5.5cm}{\includegraphics{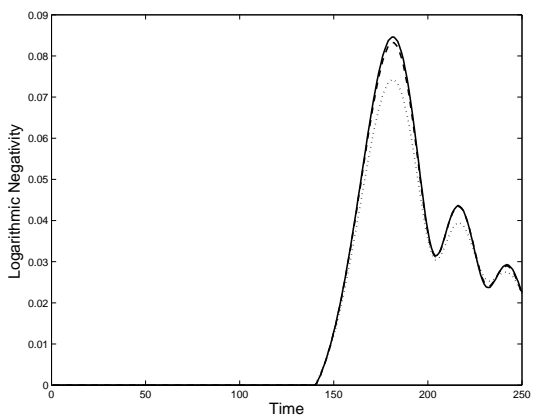}}}
\caption{\label{dopen} The same as Figure \ref{gs}, without Ohmic
dissipation and decoherence (solid line), and with Ohmic
dissipation corresponding to $Q=10000$ (dashed line) and $Q=1000$
(dotted line).}
\end{center}
\end{figure}

Decoherence mechanisms may be included without leaving the
harmonic setting. Often, the high temperature limit of Ohmic
quantum Brownian motion is appropriate with an independent heat
bath for each oscillator in the limit of negligible friction and
under the assumption of product initial conditions. Such a
decoherence mechanism can be accounted for by adding terms of the
form \cite{Giulini JKKSZ 96,Zurek 03}
\begin{equation}
-\xi [\hat q_n, [\hat q_n, \rho]]
\end{equation}
to the idealized unperturbed generator of the 
dynamical map for each of the
oscillators, where the real number $\xi$ specifies the decoherence
time scale.
%
%
However, in cases where product initial conditions are
inappropriate or unrealistic, decoherence may still be modeled
using, for example, time-convolutionless projection operator
techniques \cite{Breuer P 02}. In small systems, non-product
conditions may be incorporated by explicitly appending heat baths
to each of the oscillators with a linear coupling \cite{Zuercher T
90}, according to Hamiltonians of the form
\begin{equation}
    \hat H_n =
    \Biggl(\hat q_n \otimes \sum_{i=1}^m
    \xi_j \hat q_j^{(i)}\Biggr)
\end{equation}
with real numbers $\xi_j$, where the $\hat q_j^{(i)}$ denote the
canonical coordinates corresponding to position of the $i$-th
oscillator of the $j$-th heat bath consisting of $m$ oscillators.
Assuming a particular form of the spectral density, the coupling
strength to the finite heat baths may be chosen in a way that is
consistent with empirically known values for energy dissipation.
Often, $Q$-factors are approximately known for resonators, which
quantify the number of radians of oscillations necessary for the
energy to decrease by a factor of $1/e$. Hence, on the basis of
these $Q$-factors, the appropriate coupling may be evaluated.
Figure \ref{dopen} shows the influence of decoherence in case of
an open chain with the same parameters as in Figure \ref{gs} for
an Ohmic heat bath, i.e., $\xi_j = j \Lambda/m$, where $\Lambda>0$
is the cut-off frequency of the environment modes. One finds that
the created entanglement by suddenly switching on the interaction
is surprisingly robust against decoherence within this model.
\subsection{Entanglement transport through the harmonic chain}
In the previous section we considered the case where no
entanglement was present in the initial state of the system.
Entanglement emerged as a consequence of a sudden change in the
coupling constant between neighbouring harmonic oscillators. In
this subsection we are going to investigate a different situation
and consider the transmission of entanglement through a one
dimensional chain. To this end we initialise two harmonic
oscillators in a two mode squeezed state. We assume that one of
these oscillators is decoupled from the rest and give it the index
$0$, while the other oscillator, with index $1$, forms part of a
chain of harmonic oscillators with nearest neighbour interaction.
The remainder of the chain starts out in the ground state
corresponding to zero temperature (assuming no interactions). By
evolving the initial state, we expect the entanglement to travel
along the chain so that with increasing time more and more distant
oscillators will be entangled with the $0$-th harmonic
oscillator.
There are a number of free parameters that can be varied: the
coupling strength $c$, the amount of initial entanglement
quantified by the  two-mode squeezing parameter $r$, the time $t$
and the position of the oscillator to be entangled with the
$0$-th oscillator. In order to simplify the analytical work, we
shall be dealing with the limit $M\rightarrow\infty$.

\begin{figure}[h]
\begin{center}
\rotatebox{0}{\resizebox{!}{6cm}{\includegraphics{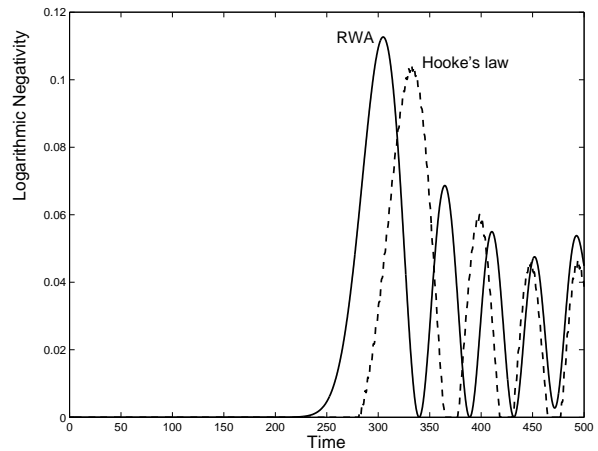}}}
\caption{\label{squeezed} The time evolution of
entanglement between the $0$-th and the $30$-{th} oscillator in
a chain with 80 oscillators with periodic boundary conditions. We
have chosen $c=0.1$ and $r=0.8$ in both cases. The entanglement
propagates slightly faster in the rotating wave interaction.}
\end{center}
\end{figure}

\begin{figure}[h]
\begin{center}
\rotatebox{0}{\resizebox{!}{6cm}{\includegraphics{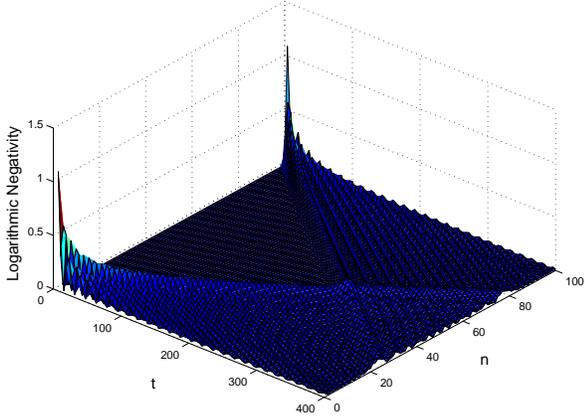}}}
\caption{\label{ringlongtime} For a ring of $40$ oscillators and a
coupling constant of $c=0.1$ the time evolution of the logarithmic
entanglement between an uncoupled oscillator and the $n$-th
oscillator is shown when initially the uncoupled oscillator and
the $1$-st oscillator are in a two-mode squeezed state with
two-mode squeezing parameter $r=0.8$. With increasing time more
and more distant oscillators are becoming entangled. Entanglement
propagates both clockwise and anti-clockwise around the ring.
After a sufficiently long time the two counter-propagating
'entanglement waves' meet at the opposite end of the ring and we
observe some entanglement enhancement.}
\end{center}
\end{figure}

While in the previous section the interaction in the rotating wave
approximation does not lead to the spontaneous creation of
entanglement, here it allows for the propagation of entanglement.
We give an example for the time evolution of the logarithmic
negativity between the $0$-th and the $30$-{th} oscillator for
both interactions in Figure \ref{squeezed}.
For both interactions we obtain qualitatively the same behaviour
but we observe that under the RWA interaction the entanglement
propagates somewhat faster but as expected this difference
decreases with decreasing coupling constant $c$. Another
difference is the fact that the entanglement under the RWA
interaction does not exhibit the small-amplitude oscillations that
the interaction due to harmonic oscillators coupled by springs
exhibits due to the existence of counter-rotating terms of the
form $\hat a_k \hat a_{k+1}$. The propagation of the quantum entanglement
can be seen even more clearly in Figure \ref{ringlongtime} where
for a ring composed of $40$ oscillators and a coupling constant of
$c=0.1$ the time evolution of the logarithmic entanglement between
an uncoupled oscillator and the $n$-th oscillator is shown when
initially the uncoupled oscillator and the $1$-st oscillator are
coupled. One observes that with increasing time more and more
distant oscillators are becoming entangled. Entanglement
propagates both clockwise and anti-clockwise around the ring.
After a sufficiently long time it becomes important that the ring
has a finite size and the two counter-rotating 'entanglement
waves' meet at the opposite end of the ring and we observe some
entanglement enhancement.
Both Figures \ref{squeezed} and \ref{ringlongtime} suggest that
entanglement can be distributed to distant oscillators. It will
therefore be interesting to study the efficiency for this transfer
when we vary the amount of entanglement provided initially by
varying $r$. In particular, we will be interested in the first
local maximum in the amount of entanglement $N_f$ as quantified by
the logarithmic negativity.
\begin{figure}[h]
\begin{center}
\rotatebox{0}{\resizebox{!}{6cm}{\includegraphics{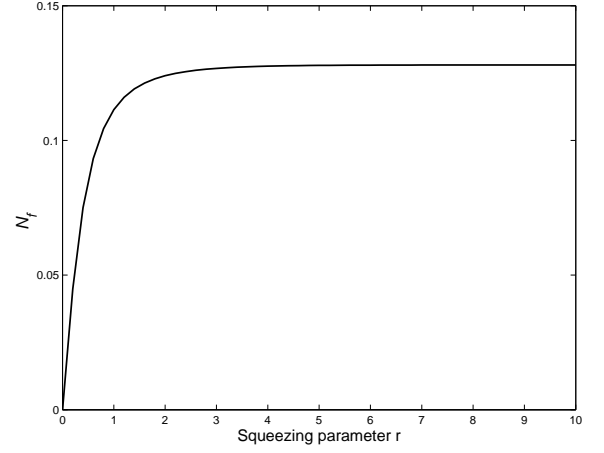}}}
\caption{\label{fig1} For oscillators interacting by springs the
graph shows the amount of entanglement at the first local maximum
as squeezing is varied. We have fixed $c=0.1$ and $n=30$. }
\end{center}
\end{figure}
We separate the study of the entanglement transfer efficiency for
the two interactions as they exhibit distinctly different
behaviours. We begin with the interaction describing oscillators
interacting via springs. Figure \ref{fig1} shows the amount of
entanglement at the first local maximum as squeezing is varied. We
observe the remarkable fact that for large initial entanglement,
the value of $N_f$ saturates. We can obtain an analytic expression
for the saturation value as a function of $c$ and $n$ by taking
the $r\rightarrow\infty$ limit. Thus, we approximate $\eta_{\rm
Spring}^2\cong -w/(8y_1)$ of Eq.\ (\ref{etaspring}) and, discarding
all the terms except $\cosh^2(r)\cong e^{2r}/4$ and
$\cosh(2r)\cong e^{2r}/2$, we obtain
\begin{equation}
    \eta_{\rm Spring}^2\cong
    \frac{\left(-2J^2_{\rm max}+(1+4c)^{-1/2}+1\right)\left(-J^2_{\rm max}+c+1\right)}{2\left(J^2_{\rm max}+1\right)^2},
\end{equation}
where $J_{\rm max}=0.6748851(n-1)^{-1/3}$ \cite{olver} is the
value of the first maximum of the $n$-th Bessel function of the
first kind. This substitution provides a very good approximation
as the first local maximum of the logarithmic negativity coincides
with that of the  $n$-th  Bessel function. Therefore we find
$N_{\rm sat}=-\log_2(|\eta_{\rm Spring}|)$ which is shown in
Figure \ref{fig2}.
\begin{figure}[h]
\begin{center}
\rotatebox{0}{\resizebox{!}{6cm}{\includegraphics{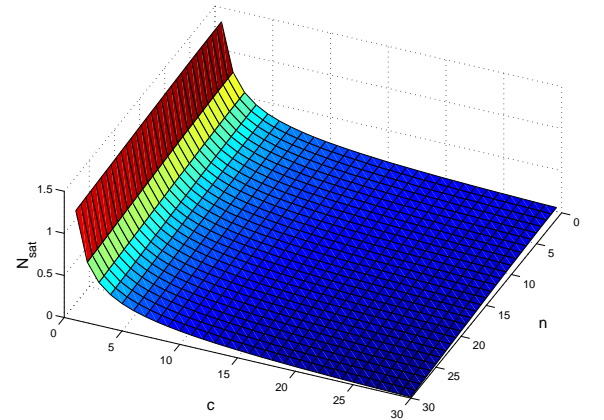}}}
\caption{\label{fig2} A graph to show how the saturation value
$N_{\rm sat}$ for the logarithmic negativity varies with the coupling
strength $c$ and the position $n$ of the second oscillator.}
\end{center}
\end{figure}
We observe that the saturation value decreases for both increased
coupling strength and increased distance. While the latter is
intuitive the former might be surprising as one could have thought
that it is advantageous to increase the coupling strength to
facilitate the transfer of entanglement but this intuition is
clearly contradicted by Figure \ref{fig2}. Even more strikingly
the entanglement vanishes entirely when the interaction strength
becomes too high. We believe that this is due to the fact that the
initial entanglement disperses across several oscillators and will
discuss this at the end of this subsection.

\begin{figure}[h]
\begin{center}
\rotatebox{0}{\resizebox{!}{6cm}{\includegraphics{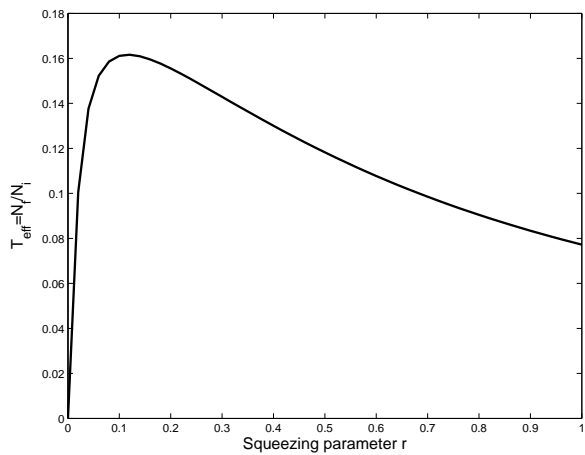}}}
\caption{\label{fig3} The efficiency of transmission as squeezing
is varied for $c=0.1$ and $n=30$. Surprisingly, the
transmission efficiency exhibits a maximum at a finite initial
entanglement.}
\end{center}
\end{figure}

If we translate these findings into an entanglement transfer
efficiency defined by
\begin{equation}
    T_{\rm eff} = \frac{N_f}{N_i},
\end{equation}
then we observe that this efficiency exhibits a non-monotonic
behaviour. Indeed, we observe a maximum in the efficiency as shown
in Figure \ref{fig3} for the same parameters as in Figure
\ref{fig1}. We have not yet found a convincing and intuitive
explanation for the occurrence of a maximum in the transfer
efficiency \cite{nonmonotonic}. In fact we will shortly see that
the phenomenon of a non-monotonic transfer efficiency is absent in
the RWA interaction. The surprising implication of this
non-monotonic behaviour of the transmission efficiency is that it
is advantageous to transmit entanglement in intermediate size
portions rather than in one very large packet.

\begin{figure}[h]
\begin{center}
\rotatebox{0}{\resizebox{!}{6cm}{\includegraphics{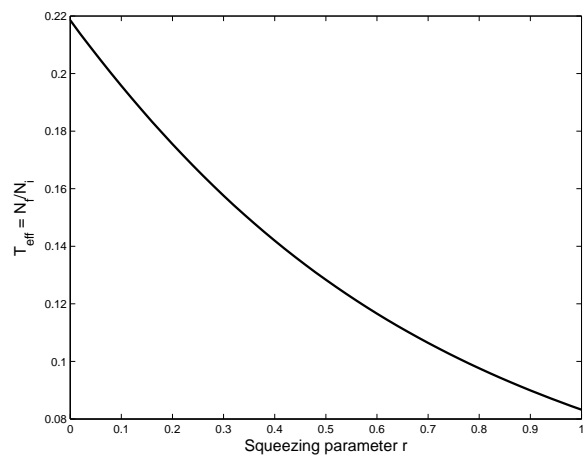}}}
\caption{\label{fig7} The efficiency of the RWA
interaction versus the two-mode squeezing parameter $r$, $n=30$
and $c=0.1$.}
\end{center}
\end{figure}

Let us now consider the same problem of the entanglement transfer
efficiency in the RWA interaction. Indeed, we can show that there
is still saturation in the amount of entanglement that can be
transmitted and we find the value of the saturation to be (after
taking $r\to\infty$)
\begin{equation}
    N_{\rm sat}=-\log_2\left(\frac{1-J_{\rm max}^2}{1+J_{\rm max}^2} \right),
\end{equation}
where again $J_{\rm max}=0.6748851(n-1)^{-1/3}$ \cite{olver} is
the value of the first maximum of the $n$-th Bessel function of
the first kind.  Note that this expression is independent of the
coupling strength $c$.  Unlike  the case of oscillators
interacting with springs there is no maximum in the efficiency for
the RWA interaction as can be seen clearly in Figure \ref{fig7}.
Indeed, for large $r$ the efficiency is tending to zero while for
$r$ approaching $0$ the efficiency tends to $T_{\rm eff} = J_{\rm
max}$.

\begin{figure}[h]
\begin{center}
\rotatebox{0}{\resizebox{!}{6cm}{\includegraphics{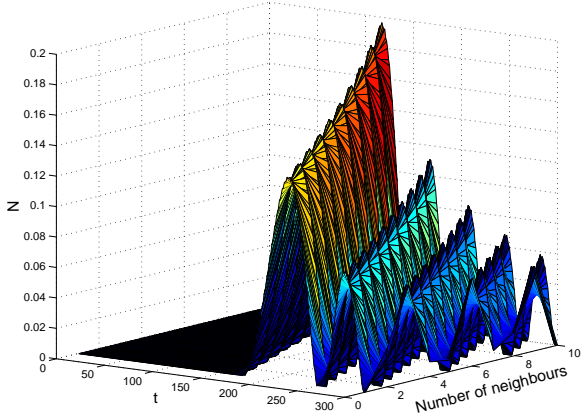}}}
\caption{\label{packet} The change in the amount
of entanglement as the number of oscillators of system $B$
centered around the $20$-{th} oscillator is increased
from $n_2=1$ to
$n_2=21$. We have a ring of 70 oscillators with squeezing
parameter $r=0.8$. Increasing the number of neighbours increases
the amount of entanglement available. This supports the idea
that the loss of entanglement is due to dispersion. }
\end{center}
\end{figure}

Since the efficiency is not equal to unity for both interactions,
the question arises as to where the rest of the entanglement is
located? The most obvious place is to search in the neighbourhood
of the $n$-th oscillator. Since we always determine the
entanglement between individual oscillators we ignore many others
that have interacted with it and have thereby become entangled.
Any entanglement between these two oscillators will therefore
deteriorate as they are being entangled with other oscillators
that we choose to ignore. This viewpoint is corroborated by
determining the entanglement between the $0$-th oscillator and a
whole group of neighbouring oscillators instead of a single one.
The result of this can be seen in Figure \ref{packet}.
This shows the change in the amount of entanglement as we
increases the number of oscillators in the second group. The graph supports 
the idea that the missing entanglement between the
$0$-th and the $n$-th oscillator is due to the creation of
entanglement between the $n$-th oscillator and its neighbours
because we start to recover entanglement as we compute the
entanglement between the $0$-th oscillator and the neighbourhood
of oscillators surrounding the $n$-th oscillator.
%
This spread of entanglement is not dissimilar to a dispersion of
the energy of a wavepacket as it experiences different group
velocities. However, the effect on the entanglement can be
considerably stronger as the energy will only decrease linearly
with the width of the wave-packet while the entanglement can drop
much more rapidly and become zero at finite spreading.

\subsection{Speed of entanglement propagation} We have found that
the propagation of half the two mode squeezed state through the
chain takes a finite time  as the entanglement between the
$0$-th and the $n$-th oscillator is exactly zero for a finite
time interval
(see for example Figure \ref{squeezed}). After a certain amount of
time, the two oscillators in question become entangled and the
logarithmic negativity reaches a temporary maximum. We are able to
determine this time analytically for  both types of interactions
that we are considering. To make the analysis tractable, we
consider an infinitely long chain. We find that the first maximum
of the logarithmic negativity coincides with the first maximum of
a Bessel function $J_m(x)$. We know the position of this maximum
occurs at $x=m+0.8086165m^{1/3}$ \cite{olver}. Noting that
$m=n-1$, $x=\zeta\Omega t=ct/\sqrt{1+2c}$ for the Hooke's law
interaction and $x=ct$ in the RWA interaction, we obtain
\begin{eqnarray}
    t_{\rm Spring} &=& \frac{n-1+0.8086165(n-1)^{1/3}}{\zeta\Omega},\label{springspeed}\\
    t_{\rm RWA} &=& \frac{n-1+0.8086165(n-1)^{1/3}}{c} \, .\nonumber
\end{eqnarray}
 We observe that the time that is required for entanglement to
be established between the $0$-th and the $n$-th oscillator is
a function of the coupling strength and the position $n$. For
large separations $n$, it quickly becomes linear in $n$ and as
expected, the larger the coupling $c$ the faster entanglement is
established. We also see that the RWA interaction produces faster
entanglement since $t_{\rm Spring}/t_{\rm RWA}=\sqrt{1+2c}$. As
$n-1$ is the separation of the $1$-st and the $n$-th oscillator
we can define the speed of propagation to be
\begin{eqnarray}
    v_{\rm Spring}&=&\frac{c}{\sqrt{1+2c}\left(1+0.8086165(n-1)^{-2/3}\right)}\cong
    \frac{c}{\sqrt{1+2c}},\nonumber \\
    v_{\rm RWA}&=&\frac{c}{1+0.8086165(n-1)^{-2/3}}\cong c. 
\end{eqnarray}
Clearly, for large $n$, the speeds approach a constant dependent
on $c$. For the RWA interaction, the propagation velocity
increases linearly with $c$. This is an attractive feature because
unlike the case of interaction via springs, the efficiency under
the rotating wave approximation does not decrease as we increase
$c$ because its efficiency is independent of $c$.

\subsection{Optimization of entanglement transfer and generation}
In the previous section we have studied the entanglement transfer
along a chain of identical harmonic oscillators as this will be
the situation that is most easily implemented experimentally.
However, we observe that the transmission efficiency decreases
with distance. One might expect that one can improve this
efficiency by tuning the couplings and the eigenfrequencies of the
harmonic oscillators suitably. Indeed, in this section we will
show what can be achieved in this more general setting. For
simplicity we consider the task of transmitting one half of a
two-mode squeezed state from one end of an open chain to the
other.

We assume as usual one decoupled harmonic oscillator with index
$0$ and a chain of length $M$ through which the other half of the
two-mode squeezed state is transmitted. Perfect transmission from
one end of the chain to the other is possible in the rotating-wave
interaction of nearest neighbours if we choose the interaction
strength
\begin{eqnarray}
    V_{n,n+1} = V_{n+1,n} = c\sqrt{n(M-n)}
\end{eqnarray}
and
\begin{eqnarray}
    V_{n,n} = 1 ,
\end{eqnarray}
with the real number $c$ being sufficiently small in order for $V$
to be positive. The choice of the diagonal elements being all
equal to $1$ is equivalent to the requirement that we choose the
eigen-frequencies $\omega_n$ of the uncoupled oscillators as
\begin{eqnarray}
    \omega_1 &=& 1 - c\sqrt{(M-1)}, \\
    \omega_n &=& 1 - c\sqrt{n(M-n)} - c\sqrt{(n-1)(M-n+1)} \, .\nonumber
\end{eqnarray}
That the transmission is perfect can be shown by first realizing
that in an interaction picture with respect to $H_0= 
\sum_i (\hat q_i^2 + \hat p_i^2)/2$ in which the diagonal elements of $V$ vanish
(this interaction picture will leave all entanglement properties
unaffected as it is of direct sum form) we can replace
\begin{equation}
  \left[\begin{array}{c}
  Q(t) \\
  P(t) \\
  \end{array}\right]=
  \exp\left({\left[\begin{array}{cc}
  0 & V_I \\
  -V_I & 0 \\
  \end{array}\right]t}\right)
  \left[\begin{array}{c}
  Q \\
  P \\
  \end{array}\right],
\end{equation}
where $Q$ and $P$ are column vectors, by the complex notation
$Q-iP$ so that
\begin{equation}
  (Q-iP)(t) = e^{iV_It} (Q-iP)\, .
\end{equation}
Now we need to realize that $V_I$ is a quantum mechanical
representation of a rotation. This allows the evaluation of the
matrix elements of $e^{iV_It}$. In particular, we find that
\begin{equation}
    (e^{iV_It})_{1M} = \left(\sin (ct/2)\right)^{M-1},
\end{equation}
so that one can generate an interchange between the first and the
$M$-th coordinate by waiting for a time $t=\pi/c$ (see Ref.\
\cite{Christandl DEL 03} for an analogous argument in spin
chains).

Without the assumption of the rotating wave approximation, i.e.,
choosing the Hamiltonian $H_{\rm Spring}$ to describe the time
evolution the above simple argument fails and indeed it is not
possible to tune the nearest neighbour couplings alone to generate
perfect transfer of entanglement for $M>2$. However, if one
chooses the couplings as above and decreases the value of the
constant $c$ then, for a fixed distance, one can obtain
arbitrarily good transfer efficiency at the expense of an
increased delay time. This should not come as a surprise, as in
the case of $c\rightarrow 0$ the rotating wave approximation
becomes exact as the terms that are neglected in the rotating wave
approximation are of order ${\cal O}(c^2)$. Therefore, for
entanglement distribution over a fixed distance they will play a
decreasing role as the time of arrival for the entanglement is of
the order ${\cal O}(c^{-1})$.

The case $M=2$ is an exception, where one may realize an exact
swap of the state of the $1$-st to the $2$-nd oscillator. To show
this, note that specific covariance matrix elements of the
$0$-th and the $2$-nd oscillator are given by
\begin{eqnarray}
    \gamma_{p_0 p_2}(t) & =
    -\sinh(r) f_1(t),\,\,\,\,
    \gamma_{q_0 q_n}(t) &=
    \sinh(r)f_1(t),\\
    \gamma_{q_0 p_2}(t) & =
    \sinh(r) \dot f_1(t),\,\,\,\,\,\,\,\,\,\,\,
    \gamma_{q_2 p_0}(t) &=
    -\sinh(r)g_1(t).\nonumber
\end{eqnarray}
Considering the functions $f_1$ and $g_1$ as specified in eqs.\
(\ref{useful1}), we find that there exists a real number $c$ and a
time $t$ such that simultaneously
\begin{equation}
    f_1(t)=1, \,\,\,\,\,\,\,\,\,\,\,
    \dot f_1(t) =0, \,\,\,\,\,\,\,\,\,\,\,
    g_1 (t) =0
\end{equation}
can be satisfied. This is the case when we chose the real $c$ such
that there exist natural numbers $k$ and $l$ such that
\begin{equation}
    c=\left( \frac{(2k +1)^2}{l^2}-1\right)/4
\end{equation}
and $t=l\pi$. Then, it follows that, as the $0$-th oscillator is
invariant and the state of the $0$-th and the $2$-nd oscillator
necessarily corresponds to a pure Gaussian state, the $1$-st
oscillator is necessarily decoupled from the other two. In this
sense the state can be swapped from one oscillator to the other,
while retaining the entanglement with the $0$-th oscillator.
Hence, one has a perfect channel for appropriate times.

\subsection{Sensitivity to random variations in the coupling}
\begin{figure}[h]
\begin{center}
\rotatebox{0}{\resizebox{!}{6cm}{\includegraphics{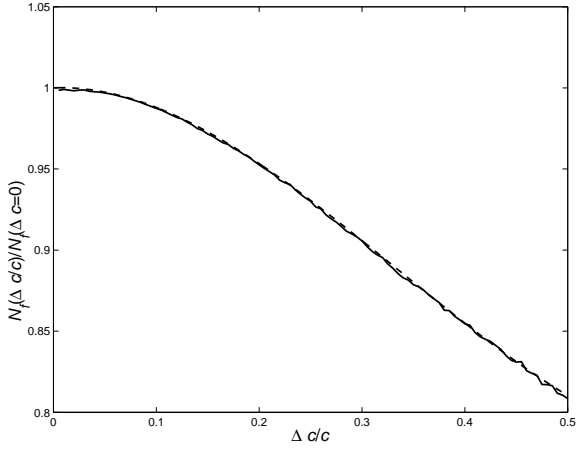}}}
\caption{\label{pertx} The ratio of the first
maximum for the case of slightly perturbed couplings over the
idealized case versus the relative perturbation size $\Delta c/c$
for a chain of length $30$, $r=0.8$ and $c=0.1$. Each data point
has been obtained as an average over 4000 realizations. The
resulting curve is very well fitted by the function $f(\Delta
c/c)= 1+0.02382 \Delta c/c - 1.60481 (\Delta c/c)^2
+1.59676 (\Delta c/c)^3$. }
\end{center}
\end{figure}

In the preceding subsections we have discussed an ideal model in
which all experimental parameters can be determined perfectly. Any
real experimental setup however will suffer small variations in
parameters such as the coupling strength between neighbouring
oscillators. In order to confirm that the effects that have been
found in this work can be observed in real experiments, we
consider in the following the impact of random position dependent
variations in the coupling strength between neighbouring
oscillators.
As an example, we consider an open chain with potential matrix
\begin{widetext}
\begin{equation}
 \hspace*{-2.5cm} V=\left[\begin{array}{ccccccc}
    1     &   0  &  0   &  0   &    \cdots   &  0 &0\\
    0     &1+c_{1,2}  & -c_{1,2}   &  0   & \cdots & 0 &0 \\
    0     &-c_{1,2}    & \ddots & \ddots   &        & &0  \\
    0     &0     & \ddots   & 1+c_{i-1,i}+c_{i,i+1} & -c_{i,i+1} &0& \vdots\\
    \vdots&\vdots&      & -c_{i,i+1} & 1+c_{i,i+1}+c_{i+1,i+2} &\ddots & 0\\
    0     &  0   &      &      &  \ddots  &\ddots  & -c_{n-1,n} \\
    0     &  0  &   0  & \cdots & 0& -c_{n-1,n}    &  1+c_{n-1,n}
\end{array}
\right],
\end{equation}
\end{widetext}
where $c_{i,j}=c+\Delta c_{i,j}$ is the position dependent
coupling between the $i$-th and $j$-th oscillators, where $\Delta
c_{i,j}$ is a realization of a random variable distributed
according to a normal $N(0,\Delta c)$ distribution. For a chain of
length $10$, an average coupling constant of $c=0.1$ and an
initial two mode squeezed state with squeezing parameter $r=0.8$.
Figure \ref{pertx} shows the ratio of the first maximum for the
case of slightly perturbed couplings over the idealized case
versus the perturbation size $\Delta c$.
We observe that for $\Delta c/c \leq 0.25$, the achieved
entanglement is greater than $95\%$ of the degree of entanglement
in the unperturbed case. Similar results apply for the RWA
interaction. These results indicate that sending quantum
information along the chain is stable under perturbations.

Similar considerations can be made for the spontaneous creation of
entanglement which show that the results are much more sensitive
to perturbations. Indeed, with the same specifications as above we
find the entanglement at the first maxima to be between a small
fraction and twice the amount for the non-perturbed case when
$\Delta c/c = 0.5$. This suggests that the experimental demands
for the verification of the spontaneous creation of entanglement
are considerably higher than for the distribution of entanglement.

\subsection{Other geometrical arrangements: Beamsplitters and interferometers}
So far we have studied only a linear chain of harmonic oscillators
through which quantum entanglement can be propagated. However, it
might be interesting to consider more complicated structures which
may be used as building blocks for more complicated networks, in
principle, any arrangement corresponding to an arbitrary weighted
graph.
\begin{figure}[b]
\begin{center}
{\resizebox{!}{4.8cm}{\includegraphics{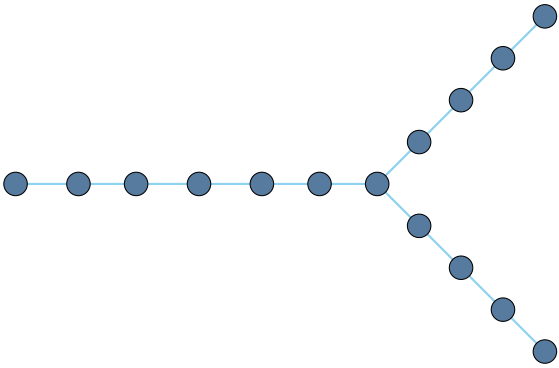}}}
\caption{\label{yshape} This figure depicts a Y-shaped structure as it is described in the text. 
A single incoming arm consists of $M_{\rm in}$ oscillators (here $7$) is
connected to two outgoing arms each consisting of $M_{\rm out}$
oscillators (here $4$).}
\end{center}
\end{figure}
In this subsection we will study briefly two possible extensions
of the linear chain, namely a Y-shaped configuration which can be
used for the generation of entanglement and a configuration
resembling an interferometer. We show furthermore how such
configurations may be switched on and off thereby controlling the
transport of quantum information in such a structure. A more
detailed discussion of such structures and their optimization will
be presented elsewhere. The material in this subsection should
merely serve as examples for possible alternative ways of creating
and manipulating entanglement through propagation in
pre-fabricated structures.

We begin by considering a chain in Y-shape which is shown in
Figure \ref{yshape}. One arm consisting of $M_{\rm in}$
oscillators is connected to two further arms each consisting of
$M_{\rm out}$ oscillators. As usual we consider nearest neighbour
interactions only and, for simplicity and the clearest
demonstration of the effects, we restrict attention to the RWA
interaction. We assume that the structure is initially in the
ground state, i.e., at temperature $T=0$. At time $t=0$ we perturb
the first harmonic oscillator exciting it either to a thermal
state characterized by covariance matrix elements
\begin{equation}
    \gamma_{q_1q_1} = \gamma_{p_1p_1} = z
\end{equation}
for some $z$, or a pure squeezed state characterized by covariance
matrix elements
\begin{equation}
    \gamma_{q_1q_1} = 1/\gamma_{p_1p_1} = z \, .
\end{equation}
\begin{figure}[b]
\begin{center}
\rotatebox{0}{\resizebox{!}{6cm}{\includegraphics{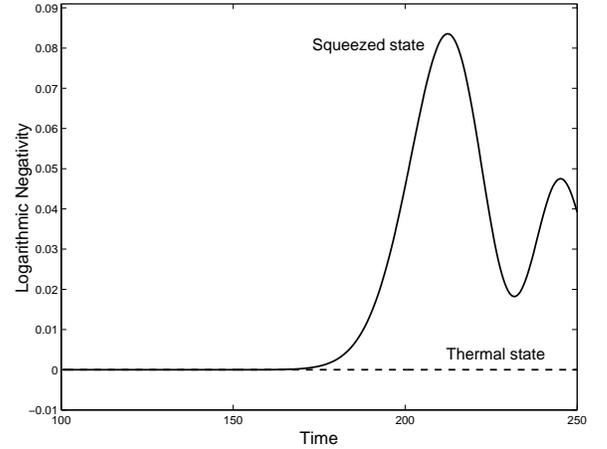}}}
\caption{\label{beamsplitter} In a chain with Y-shape and nearest
neighbour interaction of RWA type the first oscillator at the foot
of the Y-shape is either excited to a squeezed state with
$\gamma_{qq}=10=1/\gamma_{pp}$ or a thermal state with
$\gamma_{qq}=10=\gamma_{pp}$. The remaining oscillators are in the
ground state. The perturbation propagates along the chain into
both arms of the Y-shape. For an initial thermal state excitation
no entanglement is ever found between the ends of the two arms of
the Y-shape while entanglement is generated when the initial state
is a squeezed state. The coupling constant is chosen as $c=0.2$,
the arms of the Y-shape contain $30$ oscillators each while the
base contains $10$ oscillators.}
\end{center}
\end{figure}
As an example we choose a coupling constant of $c=0.2$ and let the
arms of the Y-shape contain $M_{\rm out}=30$ oscillators each
while the base contains $M_{\rm in}=10$ oscillators. In Figure
\ref{beamsplitter} we present the results for the choice of a
squeezed state with $\gamma_{q_1q_1}=10=1/\gamma_{p_1p_1}$ and a
thermal state with $\gamma_{q_1q_1}=10=\gamma_{p_1p_1}$.

We observe that for an initial thermal state excitation no
entanglement is ever found between the ends of the two arms of the
Y-shape. This can be understood because a thermal state is a
mixture of coherent states, i.e., displaced vacuum states. If the
system is initialized in the vacuum state it will evidently not
lead to any entanglement in the RWA and therefore an
initialization in a thermal state cannot yield entanglement
either. On the other hand considerable entanglement is generated
when the initial state is a squeezed state. It is possible to
optimize the generation of entanglement by adjusting the strength
of the nearest neighbour couplings but this will be pursued
elsewhere. These two observations are resembling closely optical
beamsplitters which do not create entanglement from thermal state
input but can generate entanglement from squeezed inputs (see
Ref.\ 
\cite{Wolf EP 03} for a comprehensive treatment of the entangling
capacity of linear optical devices).

\begin{figure}[h]
\begin{center}
{\resizebox{!}{3.6cm}{\includegraphics{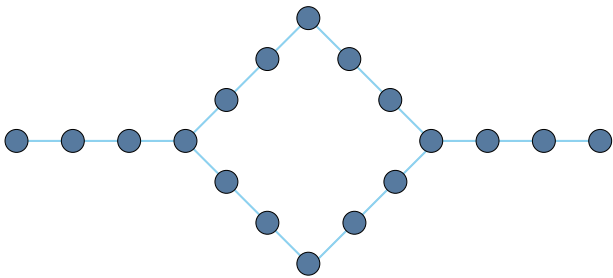}}}
\caption{\label{interferometer} A diagram of the interferometric
setup.}
\end{center}
\end{figure}

\begin{figure}[h]
\begin{center}
{\resizebox{!}{6cm}{\includegraphics{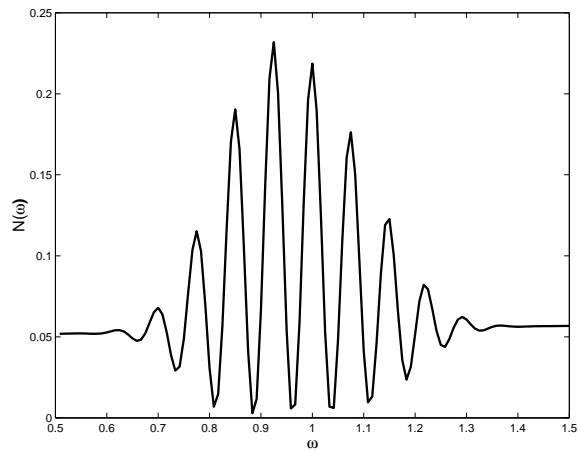}}}
\caption{\label{nfeig} The logarithmic negativity
$N(\omega)$ between the decoupled oscillator and the last one in
the interferometric configuration at the time $t=250$ plotted
against $\omega$. We have chosen $M_L=M_R-1=9$, $M_U=M_D=30$ and
$c=0.2$ in the RWA interaction. One clearly observes interference
fringes. }
\end{center}
\end{figure}

Another interesting setup is shown in Figure \ref{interferometer}.
We will henceforth call this the interferometric setup. Let the
number of oscillators on the left (including the junction), up
arm, down arm and on the right be $M_L$, $M_U$, $M_D$ and $M_R$
respectively.
If we prepare a two-mode squeezed state between a decoupled
oscillator and the leftmost oscillator of the interferometric
setup, then we are interested in how much entanglement propagates
through the setup depending on the properties of the two arms. One
may vary different parameters such as the length of one of the
arms, the coupling strength or eigenfrequency of the oscillators
in one arm. We will focus on how the change in eigenfrequency
$\omega$ of the harmonic oscillators in one of the arms affects
the propagation of entanglement through the interferometric
device. We change the eigenfrequencies of the oscillators smoothly
across one arm following
$\omega_i=1+(\omega-1) \times \min(i,M_U+1-i)/(M_U/2)$ so that the
oscillator half way through the arm has frequency $\omega$. Figure
\ref{nfeig} shows the logarithmic negativity between the decoupled
oscillator and the last one in the interferometric configuration
at the time $t=250$ plotted against $\omega$. The other parameters
are $M_U=M_D=30$, $M_L=M_R-1=9$ and $c=0.2$. One clearly observes
interference fringes in the frequency $\omega$ that are related to
the effective path-length difference between the upper and the
lower arm. The interference fringes do not have full amplitude and
their amplitude is reduced for increasing $\omega$. More
sophisticated choices for the coupling parameters in the
interferometric structure can improve on these imperfections. This
demonstrates that in interferometric structure the transmission
through the device will be strongly influenced by changes of the
properties of one arm of the structure.

This shows that more complicated structures such as the Y-shape or
the interferometric may be used to create entanglement from an
initially unentangled system and transport it. There is a distinct
analogy here to quantum optical networks which might be used for
information processing either employing the polarization degree of
freedom or as we did here the excitation number degree of freedom.
This suggests that one could construct similar 'hardwired'
networks on the level of interacting quantum systems that could
then perform certain quantum information processing or
communication tasks. This might involve structures such as the
Y-shape presented here but may also implement structures such as
interferometer structures shown in Figure \ref{interferometer} or
multi-input devices.

\begin{figure}[h]
\begin{center}
\rotatebox{0}{\resizebox{!}{6.1cm}{\includegraphics{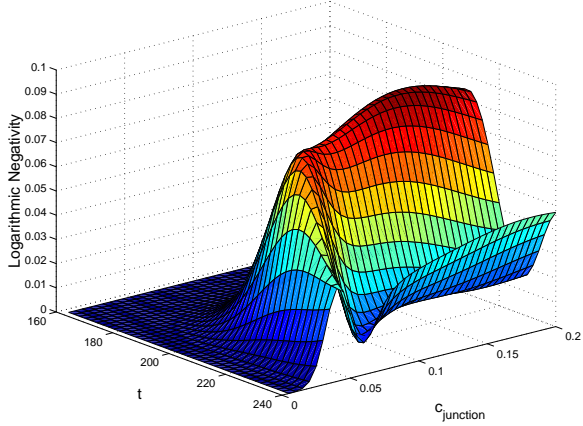}}}
\caption{\label{beamsplitterrwa_junc} The degree of entanglement in
the Y-structure in terms of the coupling strength $c_{\rm
Junction}$ of the oscillator at the three-way junction. The
parameters are chosen identically as in Figure
\ref{beamsplitter}.}
\end{center}
\end{figure}

\begin{figure}[h]
\begin{center}
\rotatebox{0}{\resizebox{!}{6.1cm}{\includegraphics{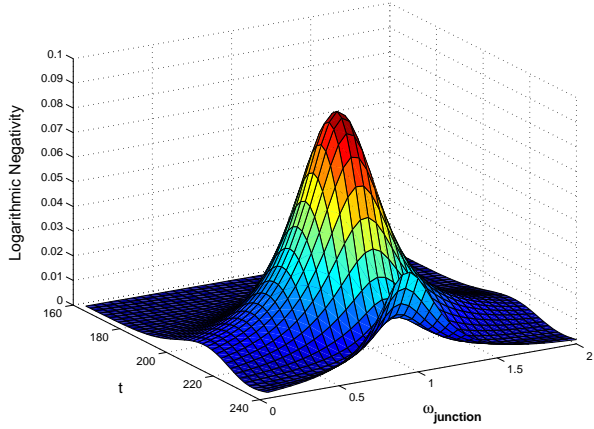}}}
\caption{\label{y_eigfreq} By increasing and decreasing
$\omega_{\rm Junction}$ respectively in the Y-shape we observe a
noticeable change in the amount of entanglement that is generated
in the device. The other parameters are chosen as in Figure
\ref{beamsplitter}.}
\end{center}
\end{figure}

If one were to consider hardwired structures, then it would be
necessary to devise methods by which these structures could be
switched on and off. Here we explore two possibilities. Firstly,
one might change the coupling strength $c_{\rm Junction}$ of the
oscillator at the three way junction in the Y-shape. Apart from
the obvious fact that they remain disentangled for $c_{\rm
Junction}=0$ (i.e., uncoupled), we find the first maximum for the
entanglement decreases to roughly half the value for large
coupling strength $c_{\rm Junction}=0.8$. This is shown in Figure
\ref{beamsplitterrwa_junc}. A further increase of the coupling
strength  to $c_{\rm Junction}=5$ does not lead to further
significant change.
A different approach would be to change the eigenfrequency or the
mass of the junction oscillator while keeping the coupling
strength the same as the other oscillators. Indeed, if we increase
the eigenfrequency $\omega_{\rm Junction}$, the entanglement can
be reduced to an arbitrarily small amount for both RWA and
spring-interactions. A decrease of the eigenfrequency is less
efficient but would also allow a significant reduction of the
amount of entanglement generated in the device. Figure
\ref{y_eigfreq} demonstrates these effects achieved by changing
the eigenfrequency of the junction oscillator in the Y-shape. It
should be noted that the dependence of the logarithmic negativity
with the eigenfrequency $\omega_{\rm Junction}$ is almost
perfectly fitted by a Lorentzian line shape.

The above examples suggest, that it is possible to switch on and
off pre-fabricated devices such as the Y-shape shown above by
adjusting either the coupling strength (decreasing it, i.e.,
approach decoupling) or the eigenfrequency (increasing it). Such a
manipulation of the junction oscillator dictates the quantum
information flowing through the junction.  An example for a
possible implementation of such a switch from an optical setting
would be coupled optical cavities which are filled with atoms.
Laser irradiation of these atoms would then lead to a shift of the
resonance frequency of the cavity  which corresponds to a change
in the eigenfrequency of a harmonic oscillator in the above
examples. In this way, individual cavities might be decoupled. A
detailed study of such a scheme will be presented elsewhere.
Finally, we would like to briefly mention an analogy with the
monotonic and non-monotonic behaviour of the efficiency. We find
that by increasing the mass of the junction oscillator in the
beamsplitter configuration, we obtain monotonically decreasing
entanglement between the two ends for the RWA interaction whereas
the Hooke's law interaction produced non-monotonic behaviour.

\section{Conclusions and Discussions}
We have investigated the entanglement dynamics of systems of
harmonic oscillators both analytically and numerically. Particular
attention has been paid to harmonic oscillators coupled by springs
(Spring) and to harmonic oscillators with a linear coupling in a
rotating wave approximation (RWA) as it is appropriate in a
quantum optical setting. After an introduction to the mathematical
formalism and the derivation of the analytical solutions for the
equations of motion for these interactions we then investigated
several possible scenarios. We considered  the generation of
entanglement without detailed local control of individual systems.
This was achieved by first switching off any interaction between
the oscillators, cooling them to near the ground state and
subsequently switching on the coupling suddenly. Surprisingly,
entanglement will be generated over very large distances which is
in stark contrast to the entanglement properties of the stationary
ground state of a harmonic chain where only nearest neighbours
exhibit entanglement \cite{Audenaert EPW 02}. We have also
demonstrated that a linear chain of harmonic oscillators is able
to transport quantum information and quantum entanglement for
various types of nearest neighbour coupling. For position
independent nearest neighbour coupling we observe that the
transmission efficiency is a non-monotonic function in the
coupling strength for Hooke's law coupling while it is
monotonically decreasing for the RWA coupling. In both cases this
suggests that it is advantageous to transmit entanglement in
smaller portions rather than large units. But due to the rapid
decline in efficiency with the spring interaction for very small
$r$, one should avoid sending in too smaller $r$. The propagation
speed for the quantum entanglement has been provided analytically.
For the above effects we have studied their sensitivity to random
variations in the coupling between the oscillators and to finite
temperatures.

Finally we have proposed more complicated geometrical structures
such as Y-shapes and interferometric setups that allow for the
generation of entanglement in pre-fabricated structures without
the need for changing any coupling constants. We have also shown
that these structures may be switched on and off by changing the
coupling of only a single harmonic oscillator with its neighbours.
This suggests the possibility for the creation of pre-fabricated
structures that may be 'programmed' by external actions. Therefore
quantum information would be manipulated through its propagation
in these pre-fabricated structures somewhat analogous to modern
micro-chips and as opposed the most presently suggested
implementations of quantum information processing where stationary
quantum bits are manipulated by a sequence of external
interventions such as laser pulses.

All these investigations were deliberately left at a device
independent level. It should nevertheless be noted that there are
many possible realizations of the above phenomena. These include
nano-mechanical oscillators \cite{Eisert PBH 03}, arrays of
coupled atom-cavity systems, photonic crystals, 
and many other
realizations of weakly coupled harmonic systems,
potentially even
vibrational modes of molecules in molecular
quantum computing \cite{Tesch R 02}. 
A forthcoming
publication will discuss device specific issues of such
realization as well as well as improved structures (including
novel topological structures as well as changes of their internal
structure such as di-atomic chains) that allow for better
performances with less experimental resources.
We hope that these ideas may lead to the development of novel ways
for the implementation of quantum information processing in which
the quantum information is manipulated by flowing through
pre-fabricated circuits that can be manipulated from outside.

We acknowledge discussion with D.\ Angelakis, S.\ Bose, D.
Browne, M.\ Christandl, M.\ Cramer, J.\ Drei{\ss}ig, 
J.J.\ Halliwell, T.\ Rudolph, M.\ Santos,
V. Vedral, S.\ Virmani, and R.\ de Vivie-Riedle. 
This work was supported by the
Engineering and Physical Sciences Research Council of the UK, the
ESF programme "Quantum Information Theory and Quantum Computing",
the EU Thematic Network QUPRODIS and the Network of Excellence
QUIPROCONE, the US Army (DAAD 19-02-0161), the EPSRC QIP-IRC, and
the Deutsche Forschungsgemeinschaft DFG. MBP is supported by a
Royal Society Leverhulme Trust Senior Research Fellowship.



\begin{thebibliography}{99}
\bibitem{nonmonotonic} Note that a similar non-monotonic behaviour
of efficiencies can be observed in different situations such as in
the efficiency of transferring entanglement from squeezed light to
two qubits \cite{Kraus C 04} or the generation of entanglement
from thermal light \cite{Plenio H 02}.
\end{thebibliography}
\end{document}